\documentclass[preprint,12pt]{elsarticle}

\usepackage[varg]{txfonts}
\usepackage{bm}
\usepackage{graphicx}
\usepackage{latexsym}
\usepackage{subfigure}
\usepackage{amssymb}
\usepackage{hyperref}
\usepackage{color}

\journal{Computer Physics Communications}

\def\rd{\mathrm{d}}

\def\cm1{\,\mathrm{cm}^{-1}}

\def\Ry{\,\mathrm{Ry}}

\begin{document}

\begin{frontmatter}

\title{ PSTGF : time-independent R-Matrix atomic electron-impact code.}

\author[a]{L. Fern\'{a}ndez-Menchero\corref{lfm}}
\author[a]{A.~C. Conroy}
\author[a]{C.~P. Ballance}
\author[b]{N.~R. Badnell}
\author[c]{D.~M. Mitnik}
\author[d]{T.~W. Gorczyca} 
\author[e]{M.~J. Seaton}

\address[a]{Centre of Theoretical Atomic, Molecular and Optical Physics.
         Queen's University Belfast.
         University Road, Belfast BT7 1NN, United Kingdom}
\address[b]{Department of Physics, University of Strathclyde. 
         107 Rottenrow East, Glasgow G4 0NG, United Kingdom}
\address[c]{Instituto de Astronom\'{\i}a y F\'{\i}sica del Espacio.
            CONICET/UBA. Buenos Aires, Argentina}
\address[d]{Department of Physics, Western Michigan University.
         Kalamazoo MI 49008-5252 USA}
\address[e]{Department of Physics and Astronomy, University College London, 
         Gower St., London WC1E 6BT, United Kingdom}
\cortext[lfm] {Corresponding author.\\
   \textit{E-mail address:} l.fernandezmenchero@qub.ac.uk}

\begin{abstract}
{\sc stgf} is a community code employed for outer-region R-matrix calculations,
describing electron-impact collisional processes.
It is widely recognised that the original version of {\sc stgf}
was written by M. J. Seaton in 1983, but through constant refinement 
over the next decades by worldwide contributors has evolved into its current form that 
more reflects modern coding practice and current computer architectures.
Despite its current wide acceptance, it was never formally published.

Therefore, we present an updated high-performance parallel version of {\sc pstgf}, 
that balances the requirements of small university clusters, yet 
can exploit the computational power of cutting edge supercomputers.
There are many improvements over the original {\sc stgf}, but most 
noticeably, the full introduction of MQDT options that provide subsequent
integration with ICFT (Intermediate Coupling Frame Transformation) codes,
and for either Breit--Pauli/DARC (Dirac Atomic R-matrix Codes), better 
load balancing, high levels of vectorisation and simplified output.
Semantically, the program is full fortran 90 in conjunction 
with MPI (Message Passing Interface) though has CUDA fortran 
options for the most numerically intensive code sections. 

\end{abstract}


\end{frontmatter}

{\bf PROGRAM SUMMARY/NEW VERSION PROGRAM SUMMARY}

\begin{small}
\noindent
{\em Program Title:} RMATRX-PSTGF  \\
{\em Licensing provisions:} GNU Lesser General Public License vLGPL2.1 \\
{\em Programming language:} Fortran90     \\

\noindent

{\em Nature of problem(approx. 50-250 words):}\\
   The R-matrix outer region code, {\sc pstgf} directly calculates various 
   electron-impact driven processes such as excitation 
   and ionisation, or provides $K$-matrices for input for subsequent
   ICFT, differential or magnetic sub-level codes. 
   As the problem size increases, there is an associated increase 
   in the input/output, the numerical computation and unbalanced workload, 
   especially in electron-impact energies where the number of open-channels 
   is equivalent to the number of closed. The code has been significantly modified 
   to address these issues. \newline
   {\sc pstgf} interfaces the R-matrix inner region with the outer region,
   with the R-matrix acting as intermediary between the two regions. The outer 
   region expresses an electron moving in the multi-pole expansion of the 
   target and predominantly employs Coulomb functions, perturbed or otherwise 
   to achieve this. This is a computationally expensive task, as the R-matrix 
   must be formed for every energy point of every partial wave. \newline
   
{\em Solution method(approx. 50-250 words):}\\
   An approach that permutes both the partial wave and energy of the incident 
   electron has been implemented.
   In this version, each processor does not calculate the same incident energy point for
   each partial wave, but rather distributes all energy points across all processors.
   This achieves better load-balancing of the work between cores and avoids the 
   case where an overloaded single processor has to always calculate in the energy 
   range where there are approximately the same number of half-open or half-closed  
   channels, which is numerically intensive. \newline
   
{\em Additional comments(approx. 50-250 words):}\\
  Dimension parameters used to define arrays and matrices within 
  {\sc pstgf} (PARAM file) have been removed, all array dimensions are dynamically
  allocatable based upon the H.DAT file and set to the exact dimension.
  A CUDA subroutine for matrix multiplications using GPUs has been included,
  it can be activated or deactivated commenting this module in the source.
  Users of serial version {\sc stgf} or older parallel versions of {\sc pstgf}
  can move to current version without any modification in the input files.
   \\

\end{small}

\section{Introduction}
\label{sec:introduction}

Time-independent R-matrix theory \cite{burke2011} is a powerful formalism that may 
be used to calculate first order collisional processes:
electron-impact excitation, ionisation, photoionisation, dielectronic recombination, 
radiative recombination for atoms and their associated ion stages.
In its simplest form, R-matrix theory splits the collisional problem 
into two distinct regions: an inner and an outer region.
The inner region is defined by the radial extent of the most diffuse 
orbital from the nucleus.
This is sufficient to encompass the charge cloud of the target, though 
mathematically this R-matrix radius may be extended further, though 
at a computational cost.
The inner region represents target atom/ion plus incident/outgoing electron, 
and is treated as $N+1$ electron many body problem, 
with all the interactions taken in account between indistinguishable particles, 
including exchange and correlation. 
We calculate a complete set of eigenfunctions representing both the bound and 
continuum spectrum.
One of the strengths of the R-matrix approach is that as we achieve a
complete description of the system in the inner region
that is independent of the incident electron energy, therefore 
only requiring one diagonalisation of $N+1$ Hamiltonian, 
for a range of electron-impact energies.
The set of codes used for the calculations in the inner region were published
in \cite{burke1992a,berrington1995,norrington1981,norrington1987} and although 
they form a foundation for this work, they are not the subject of present paper.

In the outer region, the formalism is quite different, the problem is treated as a
single electron moving in a multipole expansion of the target.
In {\sc pstgf} this single electron, which is moving in the potential created by 
the remaining electrons, is considered to be at a distance that correlation effects 
would be minimal, and therefore are not included.
Comparatively, in this region the physics simplifies, but its range may extend to 
large distances due to the long-range effect of Coulomb-like potentials. 
To extract collision strengths, we must be at distances from the nucleus where the 
wave function has returned to well-known asymptotic forms, a distance considerably 
larger than the R-matrix inner-region radius.   

The key of R-matrix method is the connection of the wave function at the boundary
between the inner and outer regions, as this interface determines the phase shifts, 
consequently the {\bf S} matrices, and finally the collision strengths.
Greater details about R-matrix method are given in~\cite{burke2011}.

{\sc pstgf} is the most common atomic outer-region program used to perform the 
calculations in the outer region for electron-impact excitation processes.
As well as directly calculating electron-impact excitation collision strengths, 
it is the progenitor code for subsequent ICFT, LS, jK and jj differential codes, 
as well as magnetic sub-level work. 
It has the capabilities to be interfaced with the molecular suite of R-matrix codes.
In general, {\sc pstgf} can be used for any process that can be described by the
time-independent and non-damped R-matrix formalism,
the most common process included in such theory is the electron-impact excitation
and de-excitation of atoms and ions; 
other important processes are electron-impact ionization, single or multiple, 
if the target structure includes such ionizing states, 
electron-impact excitation of molecules,
or any one-electron process described by time independent R-matrix theory.

{\sc pstgf} is a community code, employed by the majority of R-matrix groups 
around the world, many of whom have made contributions and improvements 
over the decades.
In present work, we build upon this to provide new sustainable version, 
that will address future problems that inevitably will be larger in scale. 

Below, are some notable additions to the code have been carried out over the years,
that due to the original code not being formally published may have been overlooked.
These include the extension to neutral atoms \cite{badnell1999c},
the ICFT (Intermediate Frame Transformation) by Griffin and 
co-workers \cite{griffin1998}, and the first  systematic parallelisation of the 
serial code by T.~W. Gorczyca \cite{gorczyca1995} and 
D.~M. Mitnik (2002, unpublished), using the MPI protocol. 
This version tried to balance the work-load by distributing the impact-electron 
energy distribution across the whole energy range for each processor, 
and remains effective when the number of processors is significantly 
less the number of energy points. 
However, in the intervening years, the number of cores available to researchers 
has increased to the point where each processor only carries out 1-3 energy points 
per partial wave.
This spurs some of the work presented later in this paper. 

Subsequently, this parallel {\sc pstgf} version was improved in efficiency.
In this new version, {\sc pstgf} was restructured in terms of its memory usage. 
At every point in the calculation, memory is assessed and if not required 
deallocated, before the next section reducing the memory footprint of the 
code by half. 
Additionally, the {\em H.DAT} input file was split into individual partial 
waves {\em H.DATXXX} files, that ensured that thousands of processors no longer 
needed to read a single file, competing against each other, 
but could concurrently read 40-100 files in smaller groups of processors.
Where possible, the LAPACK routines for matrix-matrix and matrix-vector multiplies 
where employed, especially if an optimised-vendor supplied library was available. 
Specific routines that considered the multipole perturbation of the non-perturbed 
results were heavily loop unrolled and refactored to ensure the greatest degree of 
vectorisation.
However, even these improvements need further consideration if we are to progress 
routinely to systems involving Hamiltonian matrices in excess of 
$100\,000$ by $100\,000$ and involving over $10\,000$ channels. 
In this paper, we describe how to address these problems, 
ranging from removing the last vestiges of fortran 77 legacy code 
(COMMON blocks, hard-dimension arrays), to introducing GPU enabled sections for future 
hardware compatibility. 

\section{Glossary of terms}
\label{sec:glosar}

\begin{itemize}
   \item Target state: eigenfunction of the Hamiltonian of the target, 
     $N$-electron system.
     It can be labelled with quantum numbers appropriate to the coupling scheme
     employed.
     Usually the electronic orbitals are calculated by other specialised
     packages, for example {\sc autostructure} \cite{badnell2011b},
     {\sc MCHF} \cite{froese-fischer2007}, GRASP \cite{dyal1989,parpia1996},
     or CIV3 \cite{hibbert1975}.
     R-matrix inner region codes use these orbitals to calculate the target
     energies and eigenfunctions.
   \item Partial wave $J \pi$ or $LS \pi$: symmetry of the initial system
     target state plus incoming electron, the quantum numbers are conserved.
     Each Hamiltonian representing a partial wave is calculated independently of 
     every other partial wave.
     The total cross section is then the sum of the partial cross sections for 
     all possible values of $J/L$, from zero to infinity,
     all possible couplings of $S$ (LS coupling case), and $\pi$, even and odd. \\
     For practical reasons, we have used for present work the notation of the 
     relativistic coupling $J \pi$, but the whole procedure is equivalent for the 
     non-relativistic case, and is achieved by just substituting the 
     indexes $J \pi$ for $LS \pi$.
   \item Channel: In the inner region, it is an eigenfunction of the Hamiltonian for the
     $N+1$-electron system. They are orthogonal to each other.
     The channels conserve the quantum numbers of the total angular momentum $J$
     and parity $\pi$, in the case we are working in level-resolved, relativistic 
     $J \pi$ coupling;
     or orbital angular momentum $L$, total spin $S$ and parity $\pi$ in the case
     we are working in term-resolved non relativistic $LS$ coupling.
     Hence, channels are associated to the partial waves.
     Channels can be:
   \begin{itemize}
     \item Open: a channel is classified as open when the energy of the incident 
        electron exceeds the energy between an initial term/level and a final 
        term/level. 
        In this case, a transition may be produced, the final population of the 
        channel will be larger than zero. 
        It has an oscillatory asymptotic form. 
     \item Closed: A channel is classified as closed when the incident electron energy
        is less than the energy between and initial term/level and a final term/level.
        In this case, the transition is energetically impossible. 
        It has an exponentially decaying asymptotic form. 
   \end{itemize}
   \item Phase shift $\delta_{J \pi}$: phase shift of the wave function in the 
     asymptotic region with respect to the case of a pure Coulomb potential, 
     in the case of ion target; or constant potential, in the case of neutral target.
     Transition matrices, and in consequence cross sections, may be determined in terms
     of the phase shifts.
   \item Transitions matrix $T_{if}^{J \pi}$, from initial state $i$ to final 
     state $f$: transition amplitudes. 
     Its module square $|T_{if}|^2$ is the transition probability from the initial
     target state $i$ to the final one $f$.
     The calculation of the transition matrix is particular and independent for each 
     partial wave.
     $|T_{if}|^2 = \sum_{J \pi} |T_{if}^{J \pi}|^2$ 
   \item Collision strength $\Omega(i-f)$, from initial state $i$ to final state $f$.
     Dimensionless version of the cross sections, see \cite{burke2011} for details.
     $\Omega(i-f) = \sum_{J \pi} \Omega^{J \pi}(i-f)$ 
   \item $R$-matrix $\mathbf{R}^{J \pi}$: A matrix ($n_{channels} \times n_{channels}$) 
     which connects the wave functions of the channels between the inner and the 
     outer region.
     This matrix is the key to calculate the phase shifts, and with them the 
     transition matrices, collision strengths and cross sections.
     See \cite{burke2011} for theory details.
\end{itemize}

\section{Overview}
\label{sec:overview}

{\sc pstgf} v1.1 beta 2019 is an upgrade of the previous working version 
v0.87\footnote{\url{http://connorb.freeshell.org}}.
The program is implemented in Fortran90 language, parallelised with the message 
passing interface (MPI) protocol, and with optional CUDA features for further 
optimisations if the computer architecture allows.
{\sc pstgf} reads as input the target state eigen-energies,
the channels associated with each target state for each partial wave, 
and the surface amplitudes (acquired from the matrix diagonalisation)
for each partial wave.  
All this information is encapsulated in the {\em H.DATXXX} file.
{\sc pstgf} enforces the continuity of the radial wave function and its first 
derivative between the inner and outer regions via the $\mathbf{R}$ matrix.
The main goal is the solution of the Schr\"odinger equation for 
one electron under the potential created by the multipole expansion potential of the others,
as a single particle model, neglecting the electron exchange. This may be expressed as
\begin{equation}
   \left[ - \frac{1}{2}\,\frac{\rd^2}{\rd r^2}\ +\ \frac{l\,(l+1)}{r^2}\ +\ 
   \frac{2z}{r}\ +\ \epsilon_n \right]\,F^{J \pi}_{n} =\ 
   \sum_{m=1}^{N} U^{J \pi}_{nm} F^{J \pi}_{m} \,,
\label{eq:radial}
\end{equation}
with $n$ from $1$ to $N$, number of channels in the partial wave $J \pi$,
$r$ is the radial coordinate, 
$l$ is the orbital angular momentum, which depends on the partial wave $J \pi$, 
$V=\frac{2z}{r}$ the potential in the outer region;
$z=Z-N_e$ is the effective charge.
$\epsilon_n=E-e_n$ is the channel reduced energy, 
being $E$ the impact energy of the projectile, 
and $e_n$ the excitation energy of the individual target level, 
hence $\epsilon_n > 0$ determines an open channel, 
while $\epsilon_n < 0$ a closed one. 
$F_n$ is the radial wave function,
and $U_{nm}$ is the long-range multipole expansion. 

Once equation (\ref{eq:radial}) has been solved, the radial wave function
$F_n$ is calculated from the asymptotic region to the interface between the
inner and outer regions $r_0$, then we have to match the solution with that from 
the inner region to fulfil its continuity.
This is achieved by the $\mathbf{R}$ matrix,
which performs the unitary transformation among the channels to fulfil
the continuity of the radial function and its first derivative
\begin{equation}
   F_n(r_0)\ =\ \sum_{m=1}^N R_{nm} \left(r_0\,
   \left.\frac{\rd F_m}{\rd r}\right|_{r=r_0}\ -\ bF_m \right)  \,,
\label{eq:match}
\end{equation}
$b$ is defined by the boundary conditions in the inner region, as the 
value of the logarithmic derivative of the radial wave function in the 
interface between inner and outer region. This is usually chosen to be zero.

The surface amplitudes $w_{nk}$ are defined as follows

\begin{equation}
   w_{nk}\ =\ \sum_{j=1}^{n_c} c_{njk} u_{nj}(r_0) \,,
\label{eq:wnk}
\end{equation}
where $u_{nj}(r_0)$ are the reduced surface amplitudes in the inner region; 
and $c_{njk}$ coefficients obtained from the $N+1$ Hamiltonian diagonalisation.
The summation extends to the size of the continuum basis for each orbital $n_c$, 
$n$ extends over the number of channels,
while $k$ the second index is over the Hamiltonian matrix size.
The $\mathbf{R}$ matrix itself can be defined as follows. 
\begin{equation}
   R_{nm}\ =\ \frac{1}{2 r_0}\,\sum_{k=1}^{M} \frac{w_{nk}w_{mk}}{E_k\,-\,E}\, ,
\label{eq:rmatrix}
\end{equation}
where the $E_k$ are the R-matrix poles or eigenvalues of the $N+1$ system.
This expression (\ref{eq:rmatrix}) was traditionally the part of the calculation 
which took the most of the time, but now it has been highly optimised using the 
CUDA programming techniques for GPU (Graphical Processing Units).
For further mathematical details we refer 
to \cite{burke1971,berrington1987,burke2011}.

Note in equation (\ref{eq:radial}) that the reduced energy $\epsilon_n$, 
greater or lower than zero, will determine the character of the channel, 
open or closed, so a different asymptotic behaviour, see Section \ref{sec:glosar}.
The relative number of open and closed channels for a value of the impact energy
will be relevant in terms of the computation,
we designate as $n_o$ the number of open channels 
and $n_c$ the number of closed channels.
Equation (\ref{eq:radial}) varies slightly if written in a different coupling
scheme ie. Dirac R-matrix calculations replace $l$ with $\kappa$.
Equation (\ref{eq:radial}) has to be solved for all the values of the impact energy $E$
and the complete set of partial waves $J \pi$, usually several thousands of times.
For ions, the collision strengths versus the impact energy present  
narrow Rydberg resonance structures and as a consequence the grid of the impact energies
has to be quite fine to delineate them.

Use of {\sc pstgf} assumes that the eigen-energies and eigen-functions in the 
inner region have been previously calculated, 
so we know all the channel energies $E_n$ and coefficients $w_{nk}$ for 
all the partial waves. 
This can be done with several methods and software packages;
some examples are RMATRX \cite{berrington1995},
DARC \cite{norrington1981,norrington1987},
or BSR \cite{zatsarinny2006}.
All the information about the channels in the inner region ($E_n$ and $w_{nk}$) 
is stored in a set of generic binary files {\em H.DATXXX}. 
These may be concatenated into a single {\em H.DAT} file, used by previous versions 
of the code, though for good optimisation we would advise against this. 
Ideally, the inner region has diagonalised every Hamiltonian concurrently and the 
Hamiltonians are already in this {\em H.DATXXX} form. 
{\sc pstgf} requires as input the values of the wave functions of all the channels 
at the boundary of the inner region $r_0$ for all the partial waves.
With this initial conditions, {\sc pstgf} expands the wave function in terms
of equation (\ref{eq:radial}) from $r_0$ up to a certain asymptotic limit $r_1$,
in which the wave function can be replaced by its analytic form,
of a Coulomb function in the case of a charged target,
or a spherical Bessel function in the case of the neutral target.
From this point onwards, the wave function $F^{J\pi}_n$ just follows its analytical 
asymptotic solution.
In the interface between the inner and outer region $r_0$, we have to impose the
continuity of the wave function and its first derivative.

The main calculation is distributed in two nested iterative loops.
The first loop runs over partial waves $LS \pi$ or $J \pi$, 
{\sc pstgf} has input options ({\em dstgf}) to calculate all the partial waves stored in 
{\em H.DATXXX} files or to restrict the calculation to a subset of them.
To obtain the final collision strengths $\Omega$, {\sc pstgf} has to sum up all 
the contributing partial ones $\Omega_{J \pi}$, obtained for each partial wave.
The second loop concerns the impact energies $E$, this loop runs over a set of 
$NE$ discrete values of the scaled energy $E_k$.
This grid should be fine enough to delineate fine Rydberg resonance structure, 
and therefore requires a minimum several thousand energy points.
The loop in energy is the one which is parallelised, 
the energy array $\{E_k\}$ is split among the processors $nproc$,
so each processor is assigned a set of $nproce=NE/nproc$ energies to calculate.
For an optimum performance, the number of energies should be an even divisor of $nproc$,
if this is not the case, then {\sc pstgf} will add additional points to enforce this.

The first step in the calculation is to determine which channels are open and
which ones are closed, so their asymptotic behaviour is set as boundary condition
to the equation (\ref{eq:radial}).
Then, {\sc pstgf} performs a Numerov method to propagate the coupled radial wave 
functions $F_{nm}^{J \pi}$ from the boundary of the inner region $r_0$ and the 
asymptotic limit $r_1$.
An overview of the operating mode of {\sc pstgf} is as follows: 

\begin{enumerate}
  \item Read standard input: energy grid $E$, 
        partial waves $J \pi$ to be processed, 
        other calculation parameters.
  \item Read information about the target {\em H.DAT}, $J \pi$ independent.
  \item Start loop in $J \pi$.
  \item Read information about the partial wave $J \pi$ {\em H.DAT}.
  \item Start loop in $E$. Parallel, split all the $E$ values in all the 
        processors.
  \item Calculate partial $\Omega_{J \pi}$ for each energy.
  \item End loop in $E$.
  \item End loop in $J \pi$.
  \item Add up the partial $\Omega_{J \pi}$ for all partial waves, 
        add the top-up up to $J \to \infty$, and get the total $\Omega(E)$.
  \item Write output to file {\em OMEGA}.
\end{enumerate}

\section{Computational details}
\label{sec:last}

\begin{figure}[ht]
  \includegraphics[width=0.48\textwidth]{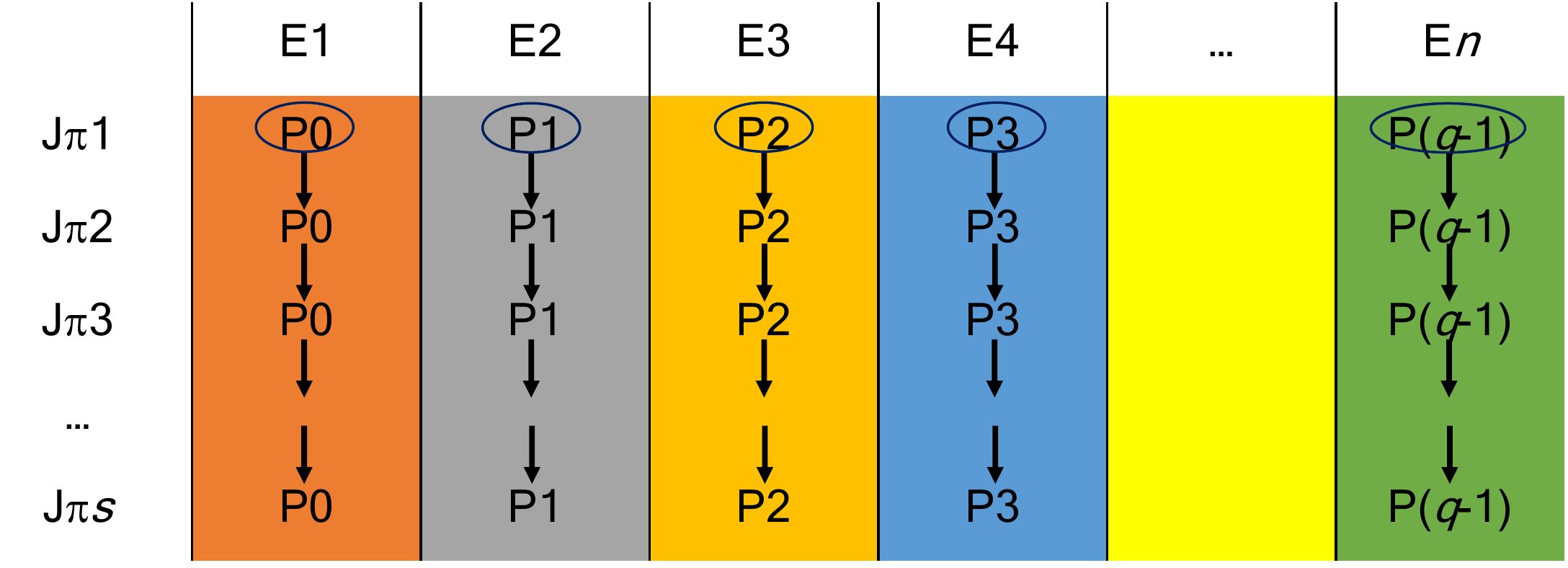} \hfill
  \includegraphics[width=0.48\textwidth]{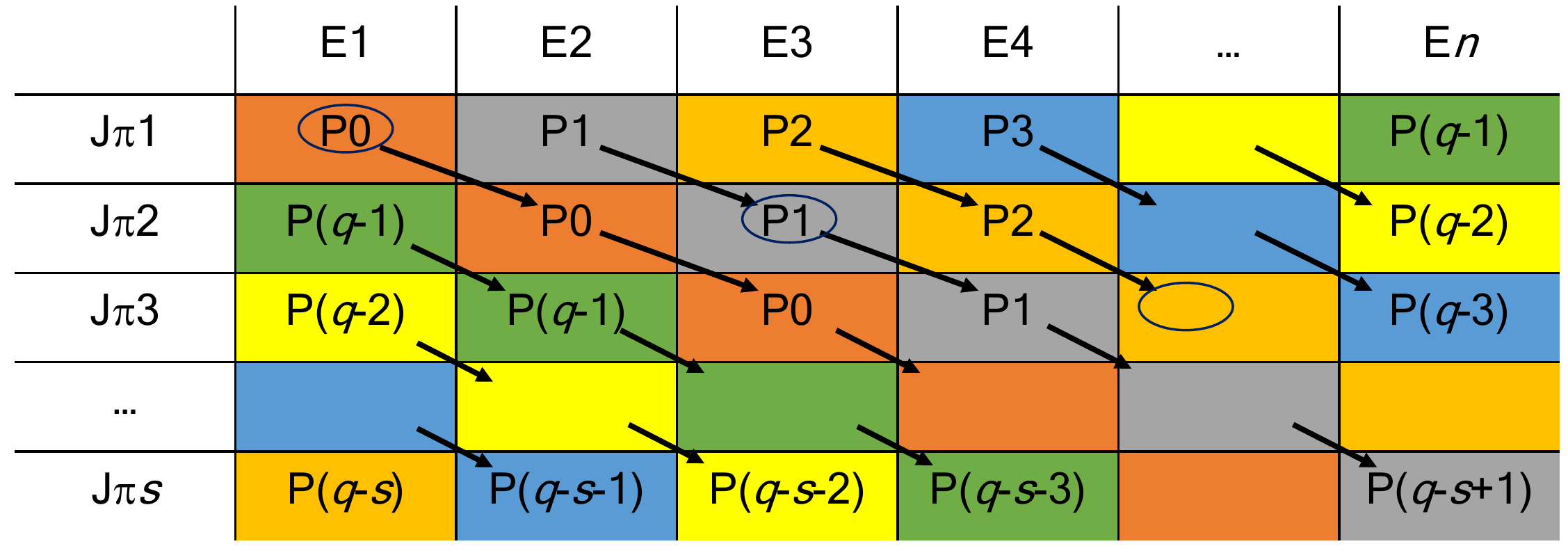}
  \caption{Colour online. Diagram of energies and symmetries worked by each processor.
     Left: v0.87; right v1.1.
     $Ei$ represents the energy array to be processed by processor $Pi$
     in the partial wave $J \pi$.
     The oval represents the partial wave were the processor $Pi$ starts 
     calculating.}
  \label{fig:ProcJE}
\end{figure}

As computer hardware capabilities have improved, more complicated systems have 
been undertaken, resulting in increasing number of symmetries and more channels 
being calculated by {\sc pstgf}, which revealed some implementation issues.
The first problem detected was the large time differences based upon different 
incident electron energies. 
The key understanding in this issue concerns the number of open and closed 
channels $n_o$, $n_c$.
It was detected that when both were approximately equal $n_o \sim n_c$, 
the computation time increases dramatically in comparison to the cases where 
either only a few channels were open, or all channels were open.
Therefore, in the input array of impact energies $E_k$, there is a distinction between 
what we call ``fast energies'' and ``slow energies''.
Usually all the partial waves include channel energies up
to a certain threshold $E_{max}$, common for all partial waves. 
Up to now, for a fixed impact energy, all the partial waves will have 
a similar number of open and closed channels.
In other words, if an energy is `fast' or `slow' for an individual partial wave, 
most probably, it will have the same character for all of them.
However, computationally the bottleneck lays in the energies around those with 
half the channels being closed and the other half open.
In previous v0.87 version, the energy array to be calculated by each processor was
independent of the partial wave, 
and all the processors worked on the same energies for each and every partial wave.
When the number of processors increases 
that gave rise to some of the processors only had a small number of slow energies, 
or even none, while other processors had to work considerably harder
for several slow energies for all the partial waves.
Hence, the time distribution among the processors was very different and
the fastest ones remained idle for large parts of the calculation, while waiting for 
the few slow ones. 
This problem is maximised when the number of processors increases, so there are
less energies per processor to proceed.
In new version v1.1, we fix this issue and improve the time balancing by having the 
energy grid different for each partial wave,
so the amount of fast energies and slow energies calculated by each processor is
evenly distributed.
Figure \ref{fig:ProcJE} shows a diagram of this energy distribution to the processors,
in the left picture (v0.87) each processor works only an energy array for all 
partial waves, while in the right picture (v1.1) each processor works all the
energies.

In v0.87 all the processors worked the same energies for all the partial waves,
and the output was split into several {\em OMEGAXXXX} files, 
$XXXX$ being the index of the processor, from $0000$ to $9999$.
Each processor was assigned a unique output file, containing unique calculated energies
which via a post-processing code, sorted and collated each file 
into a single universal {\em OMEGA} file of ascending incident energy values.
In v1.1 all processors calculate all energies, but never consecutive energies for 
the same partial wave. 
This mitigates the issue of the `slow' and `fast' energies, as no individual processor 
is assigned the half-closed/half-open channel energy for all partial waves, this 
now becomes shared among all processors. 
As consequence, each {\em OMEGAXXXX} will not contain a converged collision  
strength at each energy point, in fact each {\em OMEGAXXXX} file 
now containing every energy only has meaning when summed over the respective 
energy points in all files.
Consequently, the {\em OMEGA} output file can not be split, and therefore
the partial collision strengths are added up inside {\sc pstgf} using the 
MPI routine {\sc mpi\_reduce}, and a single {\em OMEGA} output file is produced.

A second lack of efficiency of previous versions to v0.87 was the reading of the 
input file {\em H.DAT}.
If the number of partial waves and channels increases, 
the size of the {\em H.DAT} file can reach several gigabytes.
In previous versions of {\sc pstgf}, this single file had to be read by all 
the processors each time they started the calculation for a new partial wave.
It was usual when the program was started, 
that all the processors attempted to read the same large file at the same time, 
inevitably leading to some processors blocking others, wasting computing time.
Version v0.87 had already implemented the possibility to work beyond a single 
large {\em H.DAT} file, by splitting the monolithic single {\em H.DAT}
into several smaller {\em H.DATXXX} files,
being $XXX$ the label of the partial wave, from $000$ to $999$.
Then for each partial wave, the processor must read just the individual {\em H.DATXXX} 
which contains the information about it, and not the whole {\em H.DAT} file,
with several records of irrelevant information.
Nevertheless, it was still a blocking issue when all the processors attempted to read
the same file for the same partial wave, especially at the starting of the calculation.
As mentioned above we solve this problem changing the order that the processors 
work the partial waves.
In v1.1 all the processors carry-out all the same partial waves, 
but with the difference being, not in the same order. 
This implementation avoids all the processors reading from the same file at the
same time, so these reading waiting queues are minimised.

Finally, other programming semantic improvements have been performed.
All {\sc common} blocks have been removed and replaced by {\sc module},
so the programming style is clearly {\sc fortran90}.
All the statically dimensioned arrays and matrices have been removed,
previously enforced through a parameter file ({\em PARAM}).
Now, all arrays/matrices are designated 
allocatable and their scope is dynamically assigned
based upon the values read from the {\em H.DATXXX} files. 
With this change, {\sc pstgf} does not suffer from unexpected segmentation faults,
due to a badly user defined variable in the {\em PARAM} file.
In addition, the memory usage is minimised to the essential required.
The calls to auxiliary routines from libraries {\sc lapack} and {\sc blas}
has been strategically modified in order to take advantage of their optimisation.
Another change carried out in the subroutines that contain the most load of work for 
{\sc pstgf}, is to split the memory in an strategic efficient way to work at cache level,
with faster access.

In principle, any outer-region electron-impact excitation calculation that can be
carried out with present version of {\sc pstgf} would also be possible with previous
parallel versions, or even the serial version {\sc stgf},
considering the calculation time can become huge.
The new version of the code will work without doing any changes to the input files.
Nevertheless, some minimum and optional changes can be done, and they will improve 
the efficiency of new version:

\begin{itemize}
  \item Remove the old parameter dimension information file {\em PARAM}.
        The code does not longer use it, it is absolutely redundant,
        as all the dimensions are now dynamically allocatable.
  \item The former parameter to set the maximum memory storage dimension 
        without openning any scratch file, {\sc mzmeg}, is now an input 
        variable.
        Set its value in the namelist {\sc stgf} of the standard input if you
        need to use scratch files in stead of in-memory storage 
        (default value is no scratch file to be opened).
        We hardly recommend against that unless it is really necessary,
        the hard disk can get exhausted if the number of processors is large.
  \item If the inner-region code provides one unique {\em H.DAT} file,
        split it several files {\em H.DATXXX}, each one containing one only
        partial wave.
        We recommend to use the utility tool {\sc hsplit}.
  \item Keep the partial-wave list file {\em sizeH.dat} or {\em sizeBP.dat},
        and format it properly, so the order of the partial waves in this file
        is the same than the order in the {\em H.DATXXX} files.
\end{itemize}

\section{CUDA optimisations for {\sc pstgf}}
\label{sec:cuda}

\begin{figure}[ht]
\begin{center}
  \includegraphics[width=0.7\textwidth]{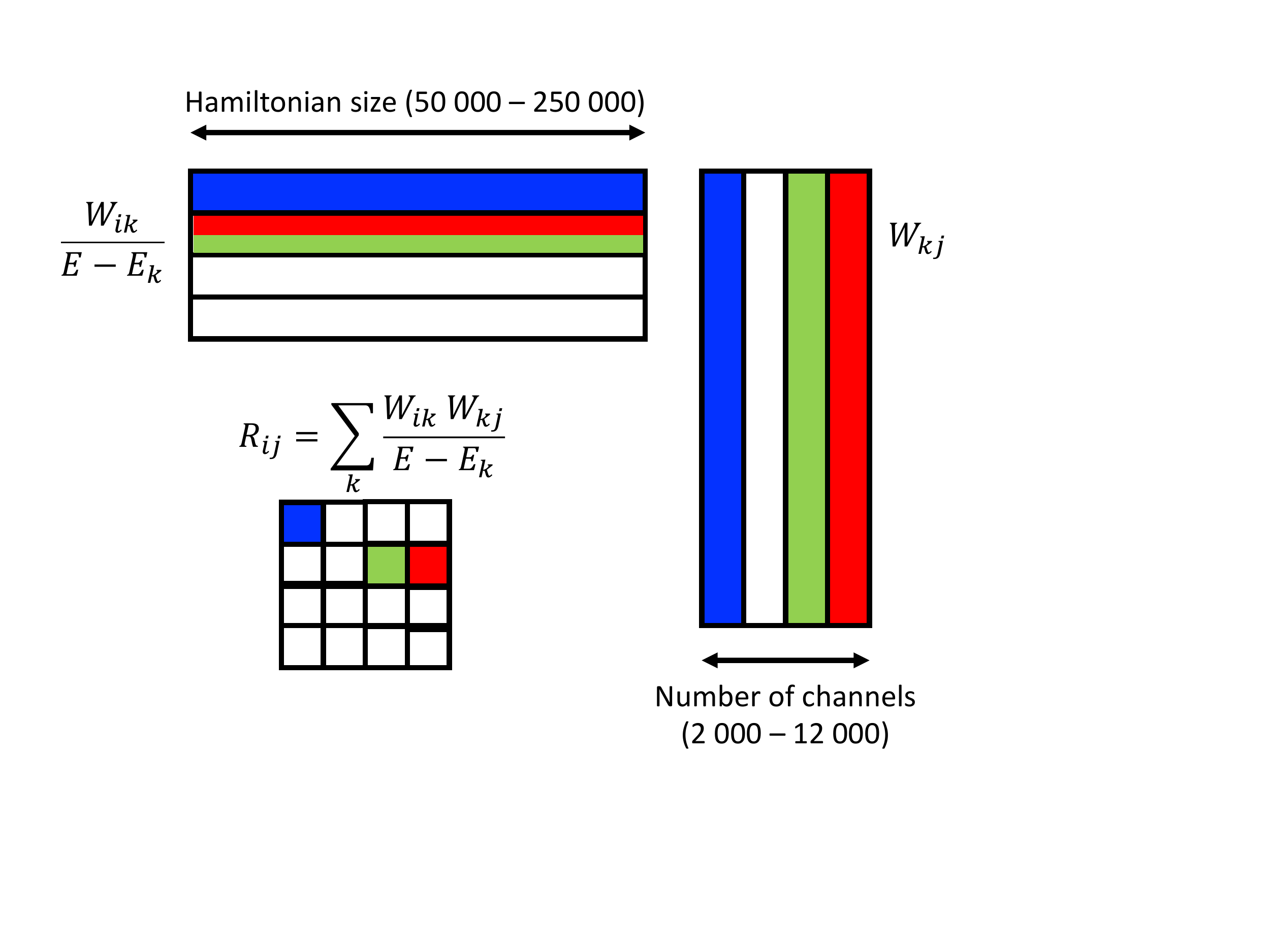} 
  \caption{Colour online. Half of the symmetric R-matrix is calculated by 10 
  matrix multiplies, and then symmetrised for the full result. 
  We optimise the memory usage of the GPU, by allowing several mpi tasks to 
  access the GPU simultaneously. 
  In the figure, the blue, red and green blocks are representative of a 
  particular matrix multiply but for different energies and partial waves, 
  and hence different R-matrices. }
  \label{fig:gpu}
\end{center}
\end{figure}

As the processor speed of individual CPUs has remained largely stagnant for 
the last decade, one option to maintain greater scalability of existing 
codes is to interface with the power of GPUs (graphical processing units).
CUDA Fortran combined with Nvidia GPUs is one of the simplest ways 
to seamlessly harness the power of MPI and the GPUs. 
GPU usage is ideally suited to dense matrix multiplies, and the initialisation 
of the R-matrix falls into this category. 
As illustrated in the Figure \ref{fig:gpu}, the R-matrix can be calculated in a 
single matrix multiplication, but this would exhaust the total memory of the GPU 
by a single processor. 
A more optimal use involves separating the R-matrix formation into ten smaller 
matrix multiplies, essentially pipelining them, one after another.
The collective result from these 10 matrix multiplies, can 
be then symmetrised to provide the full result.
This greatly reduces the consumption of memory on the GPU and allows 
more MPI tasks to concurrently access it. 
The different coloured blocks, and the respective positioning in the resulting 
R-matrix is given in Figure \ref{fig:gpu}, and corresponds to 3 different R-matrix 
matrices, at 3 different energies, being carried out concurrently.
Actual matrix multiples are further optimised by padding the larger 
matrices with zeros to ensure that the matrices fall on optimal, 
divisible by 32 or 64 boundaries.  

Traditionally, for medium to small cases the most time-consuming aspect 
of the code was the formation of the R-matrix itself (subroutine RINIT).
Test cases for neutral Fe, show that for $7261$ channels and a Hamiltonian 
size of $106\,661$ takes $14-18$ secs per R-matrix formation and achieves a factor 
of $80-100$ speed-up over existing code. 

\section{Code Structure}
\label{sec:structure}

\subsection{Input files}
\label{subsec:input}

\subsubsection{Standard input}
\label{subsubsec:standardin}

File {\em dstgf}, mandatory, ASCII. 

Structured in {\sc namelist} blocks, and optionally additional parameters.

\textbf{\textsc{namelist stgf} General calculation parameters}.

\noindent
\begin{tabular}{p{0.1\textwidth}p{0.1\textwidth}p{0.7\textwidth}}
    {\sc iprint} &
     \multicolumn{2}{p{0.8\textwidth}}{Integer from -2 to 3. Default -2} \\
     & \multicolumn{2}{p{0.8\textwidth}}{ 
      Moderates the level of standard output {\em routf}).
      Mainly used for debugging} \\
    & -2 & minimum essential output information. \\
    &  3 & maximum available output information.
           The size of the output file can become very large. \\
\end{tabular} \\
\begin{tabular}{p{0.1\textwidth}p{0.1\textwidth}p{0.7\textwidth}}
    {\sc ipert} & \multicolumn{2}{p{0.8\textwidth}}{Integer from 0 to 4. Default 0} \\
     & \multicolumn{2}{p{0.8\textwidth}}{To include the longe-range multipole potentials.
       See QDT-variable section for more details.} \\
     & 0   & Omit the long-range multipole potentials. \\
     & 1,2 & Omit long-range multipole potentials for closed channels
      when their asymptotic tail extends outside the asymptotic limit $r_1$. \\
     & 3,4 & Include all the long-range multipole potentials,
      consider the contribution from $r_1$ to infinity. \\
     & 1,3 & Perturbation for the $T$ matrix. \\
     & 2,4 & Perturbation for the $K$ matrix. \\
\end{tabular} \\
\begin{tabular}{p{0.1\textwidth}p{0.16\textwidth}p{0.64\textwidth}}
    {\sc pert} & \multicolumn{2}{p{0.8\textwidth}}{Character(3). 
      Default \texttt{"   "}.} \\
      & \multicolumn{2}{p{0.8\textwidth}}{
        Activates the perturbation, variable {\sc ipert} automatically. 
        Useful option if QDT is not activated. 
        This variable should not be included in the namelist if {\sc ipert}
        is set manually.} \\
     & \texttt{"YES"/"yes"} & 
      Activate perturbation, set {\sc ipert}$=4$. \\
     & Otherwise & Do not activate perturbation, set {\sc ipert}$=0$. \\
\end{tabular} \\
\begin{tabular}{p{0.1\textwidth}p{0.8\textwidth}}
    {\sc ac} & Float positive. Default \texttt{1.E-5}. \\
      & Accuracy/tolerance required for some numeric subroutines. \\
\end{tabular} \\
\begin{tabular}{p{0.1\textwidth}p{0.12\textwidth}p{0.68\textwidth}}
    {\sc imesh} 
    & \multicolumn{2}{p{0.8\textwidth}}{Integer, negative or from 1 to 3.
        Mandatory, no default value.} \\
    & \multicolumn{2}{p{0.8\textwidth}}{Define the energy mesh.
        See namelist {\sc imesh1}, {\sc imesh2}, {\sc imesh3} for details.} \\
    &  1 & For fixed linear grid of incremental energies. \\
    &  2 & For fixed non-linear grid based upon effective quantum number (not recommended). \\
    &  3 & Read directly the energy mesh from standard input. See additional variables section. \\
    & $-(2S+1)$ & To chose appropriate mesh for a case of spin $S$ (not recommended). \\
\end{tabular} \\
\begin{tabular}{p{0.1\textwidth}p{0.1\textwidth}p{0.7\textwidth}}
    {\sc iopt} 
    & \multicolumn{2}{p{0.8\textwidth}}{Integer, 1,2,-1,-2,10,11.
        Default 1.} \\
    & \multicolumn{2}{p{0.8\textwidth}}{Defines the partial waves $SL\pi$ or $J\pi$ 
        to be calculated.} \\
    &  1 & Calculate all the symmetries stored in the {\em H.DATXXX} files. \\
    &  2 & Specify directly in the standard input the symmetries to be calculated.
      See below format in additional variables section. \\
    & -1 / -2 & As options =1,2, but in this case treat target levels as degenerate, 
      specify the degenerate levels after namelist as additional variables, 
      see such section and {\sc nastd} parameter. \\
    & 10 / 11 & For JAJOM, obsolete and not implemented in parallel code. \\
\end{tabular} \\
\begin{tabular}{p{0.1\textwidth}p{0.8\textwidth}}
    {\sc minlt} & Integer. Default -1. \\
    {\sc maxlt} & Integer. Default 1000. \\
      & Case {\sc iopt1$=1$}, not used in case {\sc iopt1$=2$}.
      Operate only the symmetries which fulfil {\sc $L,2J\ge$ minlt} and
      {\sc $L,2J\le$ maxlt}. \\
\end{tabular} \\
\begin{tabular}{p{0.1\textwidth}p{0.1\textwidth}p{0.7\textwidth}}
    {\sc irdec} 
    & \multicolumn{2}{p{0.8\textwidth}}{Integer, 0,1,2. Default 0.} \\
    & \multicolumn{2}{p{0.8\textwidth}}{Calculate the radiative decays 
       between all the target terms or levels connected by an electric
       dipole transition.} \\
    &  0 & No radiative decays are calculated. \\
    &  1 & No QDT calculation, Bell and Seaton radiative decays. \\
    &  2 & If MQDT is activated, Hickman-Robicheaux radiative decays. \\ 
\end{tabular} \\
\begin{tabular}{p{0.1\textwidth}p{0.18\textwidth}p{0.62\textwidth}}
    {\sc lrglam} 
    & \multicolumn{2}{p{0.8\textwidth}}{Integer. Default -1.} \\
    & \multicolumn{2}{p{0.8\textwidth}}{Activate the high-$L$ top-up.
      If top-up is activated then the partial waves must be calculated in ascending
      order by all the processors, so the efficiency is reduced.
      It is recommended to use the top-up option in a single calculation including
      only the two last partial waves.} \\
    & Negative or 0 & No top-up. \\
    & Positive & Perform top-up from the highest calculated partial wave 
      up to infinity.\\
\end{tabular} \\
\begin{tabular}{p{0.1\textwidth}p{0.8\textwidth}}
    {\sc lcbe} & Integer. Default ={\sc lrglam}. \\
      & Advanced control parameter for the dipole top-up.
      Not recommended to modify its default value, only for advanced users. \\
\end{tabular} \\
\begin{tabular}{p{0.1\textwidth}p{0.1\textwidth}p{0.7\textwidth}}
    {\sc itop} 
    & \multicolumn{2}{p{0.8\textwidth}}{Integer. Default -1.} \\
    & \multicolumn{2}{p{0.8\textwidth}}{Controls the non-dipole top-up over 
      degenerate states.
      Not necessary if {\sc iopt1} is positive.
      Not recommended to modify. } \\
    & -1 & Interpolate between degenerate and non-degenerate limits when
      energy ratio exceeds $2L$. \\
    & Other & Interpolate between degenerate and non-degenerate limits when
      energy ratio exceeds the $J$ ratio. \\
\end{tabular} \\
\begin{tabular}{p{0.1\textwidth}p{0.18\textwidth}p{0.62\textwidth}}
    {\sc elas} & \multicolumn{2}{p{0.8\textwidth}}{Character(3). 
      Default \texttt{"   "}.} \\
      & \multicolumn{2}{p{0.8\textwidth}}{If the elastic collision strengths 
      should be written to output {\em OMEGA/U} file.} \\
     & \texttt{"YES"/"yes"} & Write elastic transitions to {\em OMEGA} file
       in any case. \\
     & \texttt{"NO "/"no "} & Do not write elastic transitions to {\em OMEGA} file
       in any case. \\
     & Otherwise & Write the elastic transitions to {\em OMEGA} file if the target
       is neutral, or do not if it is an ion. \\
\end{tabular} \\
\begin{tabular}{p{0.1\textwidth}p{0.1\textwidth}p{0.7\textwidth}}
    {\sc iprkm} 
    & \multicolumn{2}{p{0.8\textwidth}}{Integer from 0 to 4. Default 0.} \\
    & \multicolumn{2}{p{0.8\textwidth}}{Print $\mathbf{K}$ matrixes in optional 
      output file {\em KMAT.DAT}. For ICFT method.
      Not recommended to use this option, ICFT code is not prepared for present
      implementation of {\sc pstgf}.
      To work in ICFT formalism, use previous v0.87 version.} \\
    &  0 &  Do not write $\mathbf{K}$ matrixes. \\
    &  1 &  Each processor writes a binary, sequential file {\em KMAT.DATXXXX}
            with the physical $\mathbf{K}$ matrix elements, 
            being {\em XXXX} the processor number.\\
    &  2 &  Write unphysical $\mathbf{K}$ matrix to {\em KMATLS} file.
            Not implemented in parallel code.\\
    &  3 &  For input of post-processing differential cross sections.
            This option has been removed from code and no longer available. \\
    &  4 &  Each processor writes unphysical $\mathbf{K}$ matrixes to 
            {\em KMTLS.YYY.XXXX} files, if QDT is not activated, 
            or $S$ matrixes to {\em SMTLS.YYY.XXXX} files, if QDT is activated. 
            These files are input for ICFT code.
            $YYY$ is the number of symmetry, 
            and $XXXX$ the number of processor. \\
\end{tabular} \\
\begin{tabular}{p{0.1\textwidth}p{0.1\textwidth}p{0.7\textwidth}}
    {\sc idip} 
    & \multicolumn{2}{p{0.8\textwidth}}{Integer 0 or 1. Default 0.} \\
    & \multicolumn{2}{p{0.8\textwidth}}{Write the target dipole electric line 
       strengths} \\
    &  0 &  Do not write the line strengths \\
    &  1 &  Write the dipole line strengths $S$ to output ASCII file 
       {\em STRENGTH.DAT}. \\
\end{tabular} \\
\begin{tabular}{p{0.1\textwidth}p{0.1\textwidth}p{0.7\textwidth}}
    {\sc nomwrt} 
    & \multicolumn{2}{p{0.8\textwidth}}{Integer 0 or 1. Default \texttt{Huge(1)}.} \\
    & \multicolumn{2}{p{0.8\textwidth}}{Output management to file {\em OMEGA/U}} \\
    &   0  &  Collision strengths are not written to any file. \\
    & $>0$ &  Collision strengths $\Omega$ for first {\sc nomwrt} transitions 
      are written to output file {\em OMEGA/U} as upper triangle row-wise.  \\
    & $<0$ &  Collision strengths $\Omega$ for first {\sc -nomwrt} transitions 
      are written to output file {\em OMEGA/U} as upper triangle column-wise.  \\
\end{tabular} \\
\begin{tabular}{p{0.1\textwidth}p{0.1\textwidth}p{0.7\textwidth}}
    {\sc ibige} 
    & \multicolumn{2}{p{0.8\textwidth}}{Integer. Default 0.} \\
    & \multicolumn{2}{p{0.8\textwidth}}{Print the infinite energy limit
      collision strengths for the electric dipole transitions in 
      {\em OMEGA/U} file, flagged as negative numbers, 
      see \cite{burgess1992} for details.} \\
    & $\le 0$ &  Do not write the infinite-energy limits. \\
    & $>0$ & Write them. \\
\end{tabular} \\
\begin{tabular}{p{0.1\textwidth}p{0.1\textwidth}p{0.7\textwidth}}
    {\sc isgpt} 
    & \multicolumn{2}{p{0.8\textwidth}}{Integer. Default 0.} \\
    & \multicolumn{2}{p{0.8\textwidth}}{Write partial-wave cross sections to
      output file {\em SIGPW.DAT}} \\
    & $\le 0$ &  Do not write the partial collision strengths. \\
    & $>0$ & Write them. \\
\end{tabular} \\
\begin{tabular}{p{0.1\textwidth}p{0.8\textwidth}}
    {\sc itrmn} & Integer. Default 0. \\
    {\sc itrmx} & Integer. Default 0. \\
      & Case {\sc isgpt$=1$}, not used in case {\sc iopt1$=0$}.
        Write in output file the partial-wave cross sections from transition 
        {\sc itrmn} to transition {\sc itrmx} and also the sum of these partial 
        wave cross sections for  each $J \pi$ partial wave. \\
\end{tabular} \\
\begin{tabular}{p{0.1\textwidth}p{0.12\textwidth}p{0.68\textwidth}}
    {\sc print} & \multicolumn{2}{p{0.8\textwidth}}{Character(4). 
      Default \texttt{"FORM"}.} \\
     & \multicolumn{2}{p{0.8\textwidth}}{Print style of output file {\em OMEGA/U}} \\
     & \texttt{"FORM"} & Write the collision strengths to the final output file 
       {\em OMEGA} ASCII formatted. \\
     & Otherwise & Write the collision strengths to the final output file 
       {\em OMEGAU} binary in sequential access. \\
\end{tabular} \\
\begin{tabular}{p{0.1\textwidth}p{0.8\textwidth}}
    {\sc mzmeg} 
    & Integer. Default \texttt{Huge(1)/2**23}. \\
    & Mega-words of memory available to store in memory the {\sc omem} array,
      containing all the open-channel collision strengths for all energies.
      If the size of {\sc omem} is larger than {\sc mzmeg}$\times 8 \times 2^{20}$
      bytes, then a scratch file must be opened to store the array in hard disk.
      Warning, one scratch file is opened by each processor, 
      if the number of processor is large, the hard disk can be exhausted. \\
\end{tabular} \\
\begin{tabular}{p{0.1\textwidth}p{0.8\textwidth}}
    {\sc mzpts} 
    & Integer. Default 3201. \\
    & Number of spatial grid points of the outer region, 
      to perform the Numerov integration. \\
\end{tabular} \\

\textbf{\textsc{namelist stgf} MQDT variables}. 

To be used only if multichannel quantum-defect theory is activated in the calculation. 
This options are necessary for ICFT, otherwise it will consume a lot of computing
time unnecessarily.

\noindent
\begin{tabular}{p{0.1\textwidth}p{0.1\textwidth}p{0.7\textwidth}}
    {\sc iqdt} 
    & \multicolumn{2}{p{0.8\textwidth}}{Integer 0,1,2. Default 0.} \\
    & \multicolumn{2}{p{0.8\textwidth}}{Controls MQDT operation.
    If {\sc iqdt $>0$}, then higher dipole perturbing potentials will be included
    in terms of variable {\sc ipert}} \\
    &  0 & No QDT operation is performed. \\
    &  1 & Full MQDT, all channels treated as open, it uses unphysical 
           $\mathbf{K} / \mathbf{S}$ matrixes. \\
    &  2 & Work with unphysical $\mathbf{K}$ matrixes rather than $\mathbf{S}$ ones. \\
\end{tabular} \\
\begin{tabular}{p{0.1\textwidth}p{0.1\textwidth}p{0.7\textwidth}}
    {\sc imode} 
    & \multicolumn{2}{p{0.8\textwidth}}{Integer 0,1,-1. Default 0.} \\
    & \multicolumn{2}{p{0.8\textwidth}}{Controls read and write of unphysical 
      matrix when MQDT is activated.} \\
    &  0 & Calculates the unphysical matrix and writes it to output file JBIN. \\
    &  1 & Reads the unphysical matrix from file JBIN.
           Solution on a newly defined energy mesh obtained solely by interpolation 
           of the previous coarse mesh data read. \\
    & -1 & Single pass operation. It is performed a full solution on a 
           coarse mesh and an interpolative solution on a fine mesh. \\
\end{tabular} \\
\begin{tabular}{p{0.1\textwidth}p{0.12\textwidth}p{0.68\textwidth}}
    {\sc ijbin} 
    & \multicolumn{2}{p{0.8\textwidth}}{Integer. Default 0.} \\
    & \multicolumn{2}{p{0.8\textwidth}}{Write JBIN file in case {\sc imode=0}} \\
    &  0 & Do not write the file. \\
    & Otherwise & Write the file. \\
\end{tabular} \\
\begin{tabular}{p{0.1\textwidth}p{0.8\textwidth}}
    {\sc lmx} 
    & Integer. Default 2 (quadrupole). \\
    & If perturbations are activated, largest perturbing multipole. \\
\end{tabular} \\
\begin{tabular}{p{0.1\textwidth}p{0.1\textwidth}p{0.7\textwidth}}
    {\sc ieq} 
    & \multicolumn{2}{p{0.8\textwidth}}{Integer. Default -1.} \\
    & \multicolumn{2}{p{0.8\textwidth}}{Controls how often the unphysical 
      $\mathbf{K}$ or $\mathbf{S}$ matrix is updated.} \\
    & $<0$ & $\mathbf{K}$ or $\mathbf{S}$ matrix is updated at every 
      {\sc $|$ieq$|$}'th point of the mesh, 
      fine for constant step in a previously calculated coarser energy mesh 
      ({\sc imode}=1), not so good (inefficient) for constant step in effective 
      quantum number.  \\
    & $>0$ & $\mathbf{K}$ or $\mathbf{S}$ matrix is updated at {\sc ieq} linearly 
      spaced energies across the total energy range defined by the input energy mesh.
\end{tabular} \\
\begin{tabular}{p{0.1\textwidth}p{0.8\textwidth}}
    {\sc qetest} 
    & Float. Default \texttt{1.E-7}. \\
    & Energy in scaled Rydberg units $E/z^2$. 
      The $\mathbf{K}$ or $\mathbf{S}$ matrix is only re-interpolated when the total 
      energy has changed by more than the {\sc qetest} value since the last time.
      It gives a small time saving when using a very fine mesh. \\
\end{tabular} \\
\begin{tabular}{p{0.1\textwidth}p{0.8\textwidth}}
    {\sc fnumin} 
    & Float. Default \texttt{0.0}. \\
    & The effective quantum number below which the closed channel is omitted
      when {\sc iqdt}$=1,2$. \\
\end{tabular} \\
\begin{tabular}{p{0.1\textwidth}p{0.8\textwidth}}
    {\sc fnuhyb} 
    & Float. Default \texttt{-1.0}. \\
    & The effective quantum number below which the closed channel is a $\Theta$ 
      function rather than $S$ and $C$ ones, when {\sc iqdt}$=1,2$. \\
\end{tabular} \\

\textbf{\textsc{namelist stgf} Advanced options}

We hardly recommend against the modification of these variables, 
and they should only be worked by experienced users.

\noindent
\begin{tabular}{p{0.1\textwidth}p{0.1\textwidth}p{0.7\textwidth}}
    {\sc iomsw} 
    & \multicolumn{2}{p{0.8\textwidth}}{Integer 1,0,-1. Default 1.} \\
    & \multicolumn{2}{p{0.8\textwidth}}{Controls the number of channels to
      be treated as open in MQDT operation.} \\
    &  1 & Full MQDT: it omits closed channels with $n<l+0.1$.  \\
    &  0 & Partial MQDT: keep (with $a=-a$, {\sc iomit(ichan)}$=-1$ in 
      subroutine SC when $a<0$). \\
    & -1 & Hybrid: use $\Theta$ functions for $\nu <$ {\sc fnuhyb} and
      $\nu < l$, otherwise operate as in case {\sc iomsw}$=0$.  \\
\end{tabular} \\
\begin{tabular}{p{0.1\textwidth}p{0.8\textwidth}}
    {\sc lprtsw} 
    & Integer. Default -1 for ions and 5 for neutrals. \\
    & Value of $L$ or $2J$ above which negative {\sc ipert} is allowed. \\
\end{tabular} \\
\begin{tabular}{p{0.1\textwidth}p{0.1\textwidth}p{0.7\textwidth}}
    {\sc iccint} 
    & \multicolumn{2}{p{0.8\textwidth}}{Integer 1,0. Default 1.} \\
    & \multicolumn{2}{p{0.8\textwidth}}{Whether to include closed-closed channels 
      perturbing integrals in both MQDT or non-MQDT operations.} \\
    &  1 & Include them.  \\
    &  0 & Do not. \\
\end{tabular} \\
\begin{tabular}{p{0.1\textwidth}p{0.1\textwidth}p{0.7\textwidth}}
    {\sc intpq} 
    & \multicolumn{2}{p{0.8\textwidth}}{Integer 0,1. Default 0.} \\
    & \multicolumn{2}{p{0.8\textwidth}}{Used in internal calculation subroutine.} \\
    &  0 & use internal subroutine CORINT for closed channel MQDT $S$ 
           and $C$ integrals.  \\
    &  1 & use $\Theta$ function subroutines to generate $Q$ integrals. \\
\end{tabular} \\

\textbf{\textsc{namelist mesh1}} 

Mandatory if {\sc imesh=1}. Otherwise not present.

\noindent
\begin{tabular}{p{0.1\textwidth}p{0.8\textwidth}}
    {\sc mxe} & Integer. \\
    & Number of electron-impact energies. \\
\end{tabular} \\
\begin{tabular}{p{0.1\textwidth}p{0.8\textwidth}}
    {\sc e0} & Float. \\
    & First $z$-scaled energy of the grid in Rydberg $E/z^2$. \\
\end{tabular} \\
\begin{tabular}{p{0.1\textwidth}p{0.8\textwidth}}
    {\sc eincr} & Float. \\
    & Energy increment step. \\
\end{tabular} \\
\begin{tabular}{p{0.1\textwidth}p{0.8\textwidth}}
    {\sc qnmax} & Float. \\
    & If {\sc iqdt}$=0$ no MQDT except Gailitis average for $n>${\sc qnmax}. \\
\end{tabular} \\
\begin{tabular}{p{0.1\textwidth}p{0.1\textwidth}p{0.7\textwidth}}
    {\sc abvthr} & 
    \multicolumn{2}{p{0.8\textwidth}}{Float. Default -1.0 for ions, 
       or \texttt{1.E-3} for neutrals.} \\
    & $>0$ & Drop energies from the grid within {\sc abvthr} $z$-scaled Rydberg
      above the first excitation threshold. \\
    & $\le 0$ & Do not drop any energy. \\
\end{tabular} \\
\begin{tabular}{p{0.1\textwidth}p{0.1\textwidth}p{0.7\textwidth}}
    {\sc belthr} & 
    \multicolumn{2}{p{0.8\textwidth}}{Float. Default -1.0 for ions, 
      or \texttt{1.E-3} for neutrals.} \\
    & $>0$ & Drop energies from the grid within {\sc belthr} $z$-scaled Rydberg
      below the first excitation threshold. \\
    & $\le 0$ & Do not drop any energy. \\
\end{tabular} \\

For an optimum performance {\sc mxe} must be an even multiplier of the number of
processors.
This condition is sometimes difficult to fulfil if {\sc abvthr} or {\sc belthr}
are used, so energies are dropped from the grid.
This happens commonly in the case of neutrals.

\textbf{\textsc{namelist mesh2}} 

Mandatory if {\sc imesh=2} or negative, otherwise not present. 
This option is not yet implemented in parallel version {\sc pstgf}.

\textbf{\textsc{namelist mesh3}}

Mandatory if {\sc imesh=3}, otherwise not present.

\noindent
\begin{tabular}{p{0.1\textwidth}p{0.8\textwidth}}
    {\sc mxe} & Integer. \\
    & Number of electron-impact energies. 
      To be read afterwards as additional parmeter. \\
\end{tabular} \\

Additional variables not sorted as namelist, mandatory for determined values of 
the variables in namelist {\sc stgf}, see above.

\textbf{Case \textsc{iopt1}=2}.

In this case, the partial waves to be calculated must be specified.
It is expected several lines, each one with three integer numbers S, L, PI.

\noindent
\begin{tabular}{p{0.14\textwidth}p{0.1\textwidth}p{0.66\textwidth}}
    S, L, PI & 
    \multicolumn{2}{p{0.8\textwidth}}{Several lines of three free-format integers each} \\
    \ldots & 
    \multicolumn{2}{p{0.8\textwidth}}{List of partial waves to be calculated.
      These symmetries should be present in files {\em H.DATXXX}.} \\
    & S  & $2S+1$ of the partial wave, or $0$ to flag $J \pi$ coupling. \\
    & L  & $L$ or $2J$ of the partial wave. \\
    & PI & Parity of the partial wave: 0 for even, 1 for odd. \\
    \texttt{-1 -1 -1} & 
    \multicolumn{2}{p{0.8\textwidth}}{or EOF marks end of list.} \\
\end{tabular} \\

\textbf{Case \textsc{iopt1}=-1,-2}.

In this case, the target term or level energies are treated as degenerate, 
and the number of degenerate terms or levels for each target energy must be read in.
It is expected a first line with an integer number {\sc nastd}, and a second
line with a set of {\sc nastd} integers {\sc nlev(1:nastd)}.
In the case {\sc iopt1}$=-2$ the list of partial-wave symmetries is specified
after the shortlist of the degeneration of the energies.

\noindent
\begin{tabular}{p{0.1\textwidth}p{0.8\textwidth}}
    {\sc nastd} & Integer, free format. \\
    & Number of target degenerated energies. \\
\end{tabular} \\
\begin{tabular}{p{0.16\textwidth}p{0.64\textwidth}}
    {\sc nlev(1:nastd)} & {\sc nastd} integers, free format. \\
    & Number of degenerated terms or levels for each energy. \\
\end{tabular} \\

\textbf{Case \textsc{imesh}=3}.

After the namelist {\sc mesh3}, expected a total of {\sc mxe} floating 
point numbers.

\noindent
\begin{tabular}{p{0.16\textwidth}p{0.64\textwidth}}
    {\sc emesh(1:mxe)} & {\sc mxe} float, free format. \\
    & Electron-impact $z$-scaled energies in Rydberg, 
      sorted from smallest to largest.
      For an optimum performance, {\sc mxe} should be an even multiplier of the
      number of processors.  \\
\end{tabular} \\

\subsubsection{H.DAT}
\label{subsubsec:hdat}

Mandatory, binary.

Binary output from the inner region codes, used as input for the outer region codes.

It can be presented in three different ways.
{\sc pstgf} inquires which one is present in the following order,
if several types are present, just the first one inquired as positive is
used, the remaining ones are just ignored:

\begin{enumerate}
  \item Several {\em H.DATXXX} files, starting by {\em H.DAT000} (recommended).
    {\em H.DAT000} contains the information about the target and at least the first
    $J \pi$ or $LS \pi$ partial wave.
    Each {\em H.DATXXX} file contains information about the channels of 
    one (recommended) or several partial waves.
    In the case the number of {\em H.DATXXX} files is the same than the number
    of lines in the {\em sizeH.dat} file and the sorting of the {\em XXX}
    indexes agrees with the sorting of partial waves in {\em sizeH.dat} 
    (see \ref{subsubsec:sizehdat}) the performance will be optimum.
  \item One single {\em H.DAT} file.
    The single file contains the information about the target and the channels of 
    all the partial waves calculated in the inner region.
    {\sc pstgf} will still work, but its performance will be not optimum due to
    all the processors will have to read the same large file for all the 
    symmetries.
  \item One single {\em DSTGH.DAT} file.
    For inner region calculations with the old version DARC code, now obsolete.
    The option is kept just to allow the code work for older calculations, but
    it is not recommended for new work.
\end{enumerate}

\noindent
\begin{tabular}{@{}r@{:\ }l|l}
  \hline
  \multicolumn{3}{l}{H.DAT000, Target information} \\
  \hline
  1 & 5 INT*4, 2 REAL*8 & \texttt{NELC,NZ,LRANG2,LAMAX,NAST,RA,BSTO} \\
  2 & \texttt{NAST} REAL*8 & \texttt{(ENAT(I) \quad I=1-NAST)} \\
  3 & \texttt{NAST} INT*4   & \texttt{(LAT(I) \quad I=1-NAST)} \\
  4 & \texttt{NAST} INT*4   & \texttt{(ISAT(I) \quad I=1-NAST)} \\
  5 & 3\,\texttt{LRANG2} REAL*8 & \texttt{((COEFF(I,L) \quad I=1-3), L=1-LRANG2)} \\
  \multicolumn{2}{l|}{\texttt{L=1-LRANG2}} & \\
  \texttt{L1} & 1 INT*4 & \texttt{NBUTD(L)} \\
  \texttt{L2} & \texttt{NBUTD} REAL*8 & \texttt{EBUTD(I,L) \quad I=1-NBUTD} \\
  \texttt{L3} & \texttt{NBUTD} REAL*8 & \texttt{CBUTD(I,L) \quad I=1-NBUTD} \\
  \hline
  \multicolumn{3}{l}{H.DAT000, First symmetry information} \\
  \hline
  1 & 6 INT*4 & \texttt{LRGL2,NSPN2,NPTY2,NCHAN,MNP2,MORE2} \\
  2 & \texttt{NAST} INT*4 & \texttt{(NCONAT(I) \quad I=1-NAST)} \\
  3 & \texttt{NCHAN} INT*4 & \texttt{(L2P(I) \quad I=1-NCHAN)} \\
  4 & \texttt{LAMAX$\times$NCHAN$^2$} REAL*8 & \texttt{(((CF(I,N,M) \quad I=1-NCHAN),N=1-NCHAN),M=1-LAMAX)} \\
  5 & \texttt{MNP2} REAL*8 & \texttt{(VALUE(I) \quad I=1-MNP2)} \\
  6 & \texttt{NCHAN$\times$MNP2} REAL*8 & \texttt{((WMAT(I,K) \quad K=1-NCHAN),I=1-MNP2)} \\
  \hline
\end{tabular}

Information about the target, first set of records of {\em H.DAT000} or 
{\em H.DAT} files:

\begin{itemize}
  \item 1: Number of electrons; nuclear charge; maximum $L$ or $2J$ (target); 
     maximum calculated coupling multipole; number of target terms or levels;
     R-matrix box size (a.u.); logarithmic derivative of the radial function 
     in the boundary.
  \item 2: Target level or term energies.
  \item 3: Target $L$ or $2J$ of each term or level.
  \item 4: Target $2S+1$ of each term, or 0 if $J \pi$ coupling.
  \item 5: Coefficients of expansion of the target wave functions for each
     term or level, see \cite{burke2011} for details.
  \item Following 3$\times${\sc lrang2} records: coefficients of the Buttle correction.
     For older versions of inner-region codes, these data could be in the different 
     file {\em DBUT} in stead of {\em H.DAT(000)}.
\end{itemize}

Information about the $N+1$-electron symmetries, following records of {\em H.DAT}
or {\em H.DAT000} files, after the target information;
or whole files {\em H.DATXXX} files, with $XXX \ne 000$.

\begin{itemize}
  \item 1: $L$ or $2J$; $2S+1$ or $0$; parity of partial wave;
    number of channels; number of Hamiltonian eigenvalues in partial wave;
    flag to check if present partial wave is the last one to be read.
  \item 2: Number of channels attached to the target level $I$.
  \item 3: $L$ or $2J$ of the target term or level associated to channel $I$.
  \item 4: Coefficient of the multipole potential expansion.
  \item 5: Eigenvalues of the Hamiltonian, R-Matrix pole energies $E_k$
     equation \ref{eq:rmatrix}.
  \item 6: Eigenvectors of the Hamiltonian, R-Matrix amplitudes $w_{ij}$ 
     equation \ref{eq:rmatrix}.
\end{itemize}

\subsubsection{sizeH.dat}
\label{subsubsec:sizehdat}

Optional, nevertheless highly recommended, ASCII.

All the codes that work the inner region and provide {\em H.DAT} files
are able to provide an auxiliary formatted {\em sizeH.dat} file.
There are two kinds of {\em sizeH.dat} files readable by {\sc pstgf}.
It inquires if they exist in the following order:

\begin{enumerate}
  \item {\em sizeH.dat} for $LS \pi$ or $J \pi$ coupling.
  \item {\em sizeBP.dat} for $J \pi$ coupling only.
\end{enumerate}

If both files exist, only {\em sizeH.dat} is read by {\sc pstgf} and 
{\em sizeBP.dat} is ignored.
{\em sizeH.dat} contains the basic information about each partial wave:
$2S+1$, $L$, $\pi$, number of channels, number of continuum basis $\times$ number of channels, 
number of eigenvalues for the $N+1$ Hamiltonian.
If $2S+1=0$ then {\sc pstgf} assumes that partial waves are presented in
$J \pi$ coupling and $J=2L$.
The number of lines in the file is interpreted by {\sc pstgf} as the number of partial
waves stored in the {\em H.DATXXX} files (see \ref{subsubsec:hdat}).
The best performance is reached if there is exactly one partial wave per
{\em H.DATXXX} file and the sorting of the {\em XXX} indexes agrees with the sorting 
of partial waves in {\em sizeH.dat} 

{\em sizeBP.dat} file works the same way as {\em sizeH.dat} file but it has
no value of $S$ stored, it assumes always $J \pi$ coupling.

If no {\em sizeH.dat} or {\em sizeBP.dat} files are found or they have different
number of lines than the number of {\em H.DATXXX} files,
then a preliminary read of the {\em H.DATXXX} files has to be carried out in order
to let the program know in which file is stored which partial wave, 
and that will take some time if the files are large.
A correctly formatted {\em sizeH.dat} will help a good performance of {\sc pstgf}.

\noindent
\begin{tabular}{@{}r@{:\ }l|l}
  \hline
  \multicolumn{3}{l}{sizeH.dat} \\
  \hline
  \texttt{1-NSLPI} & \texttt{8X,I5,10X,I6,8X,I6,3(4X,I3)} & 
  \texttt{NCHAN,NCON,MNP2,S,L,IPI} \\
  \hline
  \multicolumn{3}{l}{sizeBP.dat} \\
  \hline
  \texttt{1-NSLPI} & \texttt{3I7,9X,I3,8X,I3} & \texttt{NCHAN,NCON,MNP2,J2,IPI} \\
  \hline
\end{tabular}

The file is ASCII formatted and has {\sc nslpi} lines, number of partial waves. 
In each partial wave it is read the number of channels; 
the number of continuum functions ({\sc nchan}$\times${\sc nrang});
number of eigenvalues of the Hamiltonian ({\sc ncon} plus bound functions);
$2S+1$ or $0$; $L$ or $2J$; parity.

\subsubsection{DBUT}

File to store the Buttle correction.
It works only for old versions of inner-region codes, currently the Buttle correction
is integrated in {\em H.DATXXX} files.

%

\subsection{Output files}
\label{subsec:output}

\subsubsection{Standard output}
\label{subsubsec:standardout}

File {\em routf} from processor 0 and {\em routfgXXXX} from other processors, 
being $XXXX$ the processor number, ASCII. 
Standard output, execution information, warnings and errors.
If the calculation finished OK, with no errors, then the standard output files from 
processors different to 0 will have no additional information, 
and they can be removed to save space in the hard disk.

\subsubsection{OMEGA/U}
\label{subsubsec:omega}

Final results, binary or ASCII depending on input variable {\sc print}.

One single file, if in the standard input the variable {\sc print} is specified
as ``FORM'' then output will be formatted in an {\em OMEGA} file,
any other value and it will be unformatted with sequential access 
in an {\em OMEGAU} file.

\noindent
\begin{tabular}{@{}r@{:\ }l|l}
  \hline
  \multicolumn{3}{l}{OMEGA(U)} \\
  \hline
  1 & 2 INT*4 & \texttt{NZ \quad NEL} \\
  2 & 3 INT*4 & \texttt{NLEV \quad MXE \quad NOMWRT} \\
  3 & 2$\times$\texttt{NLEV} INT*4 & \texttt{(S(I) \quad L(I), \quad I=1-NLEV)} \\
  4 & \texttt{NLEV} REAL*8 & \texttt{(ELEV(I), \quad I=1-NLEV)} \\
  \texttt{J=1-MXE} & \texttt{NOMWRT}+1 REAL*8 & \texttt{E(J), (OMEGA(J,I), \quad I=1-NOMWRT)} \\
  \hline
\end{tabular}

In any of the formats it contain the following information:

\begin{itemize}
  \item 1: nuclear charge and number of electrons of the target.
  \item 2: Number of target terms or levels, 
    number of impact energies and number of transitions stored, 
    {\sc nomwrt}, or between all the target levels, it can also contain
    elastic transitions if variable {\sc elas} is set to \texttt{"YES"}.
  \item 3: Array containing the $L,S$ or $0,2J$ of each 
    target term or level.
  \item 4: Array containing the term or level excitation energies of the 
    target in Rydberg scaled units with respect to the ground state energy.
  \item $J=1-MXE$: Impact scaled energy in Rydberg $E/z^2$ and 
    array with the $\Omega$ values for the collision strengths for that energy.
    The $\Omega(i-j)$ matrix is stored as the upper triangle by rows or columns
    depending of the sign of the variable {\sc nomwrt}.
\end{itemize}

\subsubsection{SIGPW.DATXXXX}

$XXXX$: index of processor.

ASCII, written if {\sc isgpt}$=1$.

Partial wave transition amplitudes, to calculate partial cross sections.

\subsubsection{JBINLS.DATXXXX}

$XXXX$: index of processor.

Binary, written if {\sc iprkm}$=4$.

Channel information to input in STGICF.

\subsubsection{SMTLS.YYY.XXXX}

$XXXX$: index of processor;
$YYY$: index of symmetry.

Binary, written if {\sc iprkm}$=4$ and MQDT active.

Unphysical $\mathbf{S}$ matrix, to be used as input for STGICF.

\subsubsection{KMTLS.YYY.XXXX}

$XXXX$: index of processor;
$YYY$: index of symmetry.

Binary, written if {\sc iprkm}$=4$ and MQDT active.

Unphysical $\mathbf{K}$ matrix, to be used as input for STGICF.

These two files should be used with care, as the output is sorted by each processor
in terms of its local energy grid.
ICFT code does not work with present implementation of {\sc pstgf},
so these files can not be used unless {\sc pstgicf} is updated in a consistent way
of current {\sc pstgf} version.

\subsubsection{OMEGDR}

ASCII, written if {\sc ndrmet}$>0$. 
Same format of {\em OMEGA} file, replacing the number of transitions 
{\sc nomwrt} for initial states for DR {\sc nast}.

Dielectronic recombination cross sections. 
{\sc pstgf} is not the appropriate code to use to calculate DR. 
The radiation-damped code {\sc pstgfdamp} should be used instead.

\subsection{STRENGTH.DAT}

ASCII, written if {\sc idip}$=1$.

E1 line strengths $S$ between all target terms or levels.

%

\subsubsection{TERM.DAT}

ASCII, written if {\sc iprkm}$=4$.

Information about the target terms or level: $2S+1$, $L$, $\pi$ and $E$.

\subsection{Scratch files}

\subsubsection{SCRATCH1}

File to store the {\sc omem} array if its dimension is larger than 
{\sc mzmeg$\times 2^{20}$}.
In this case a scratch file is opened and the array {\sc omem} is stored on the hard disk.
One file is opened by each processor, if the number of processors is large, 
caution must be taken to ensure that hard disk is not unexpectedly filled.

The best practice is not to modify the default value of {\sc mzmeg} to ensure that no scratch
files are open.
There is also a reduction the input and output time.

\section{Test cases}
\label{sec:calculations}

\begin{figure}
  \includegraphics[width=\textwidth]{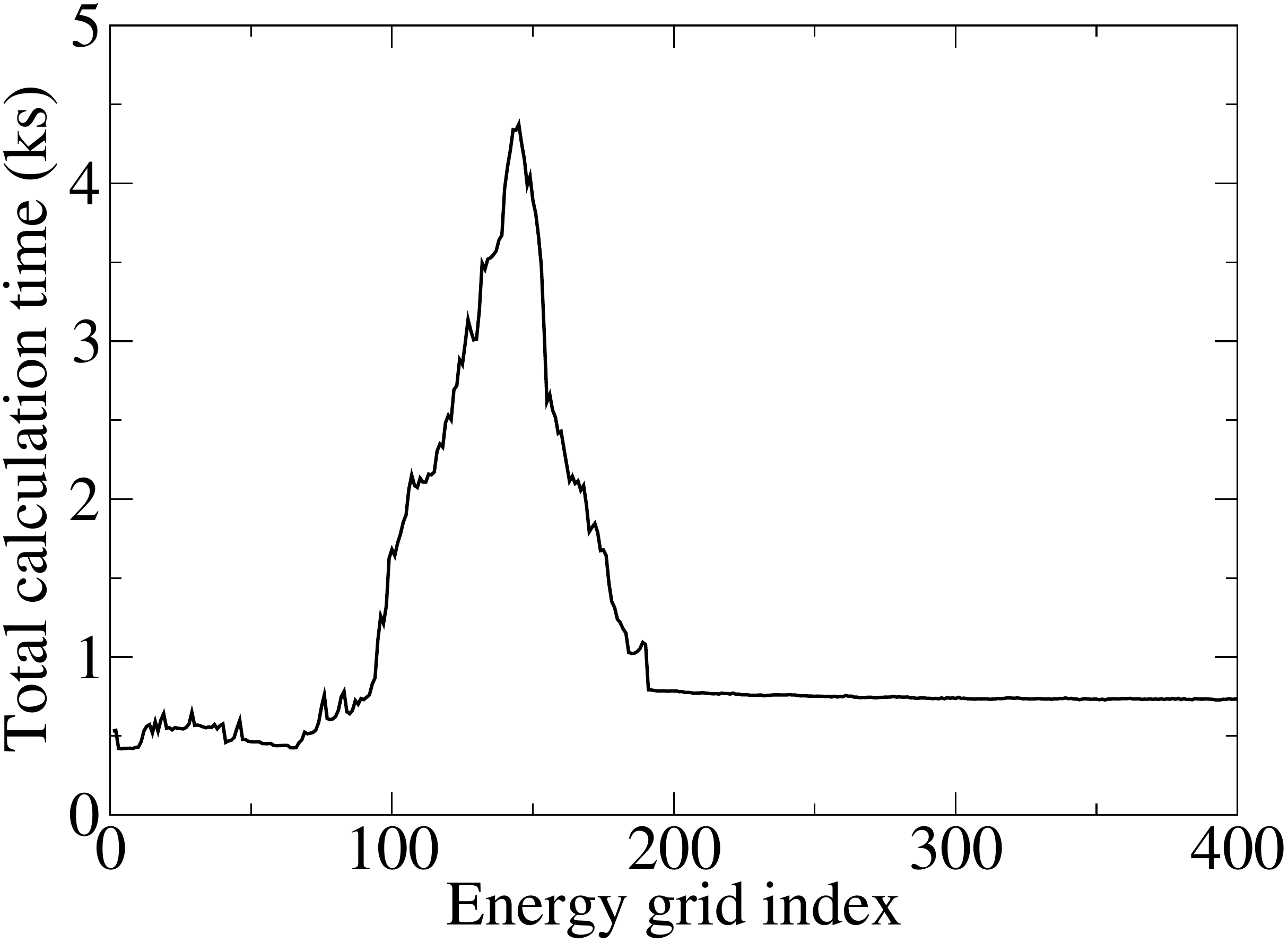}
  \caption{Total processor time spent in each energy $\mathrm{Ni^{3+}}$}
  \label{fig:timeenNi3}
\end{figure}

\begin{figure}
  \includegraphics[width=\textwidth]{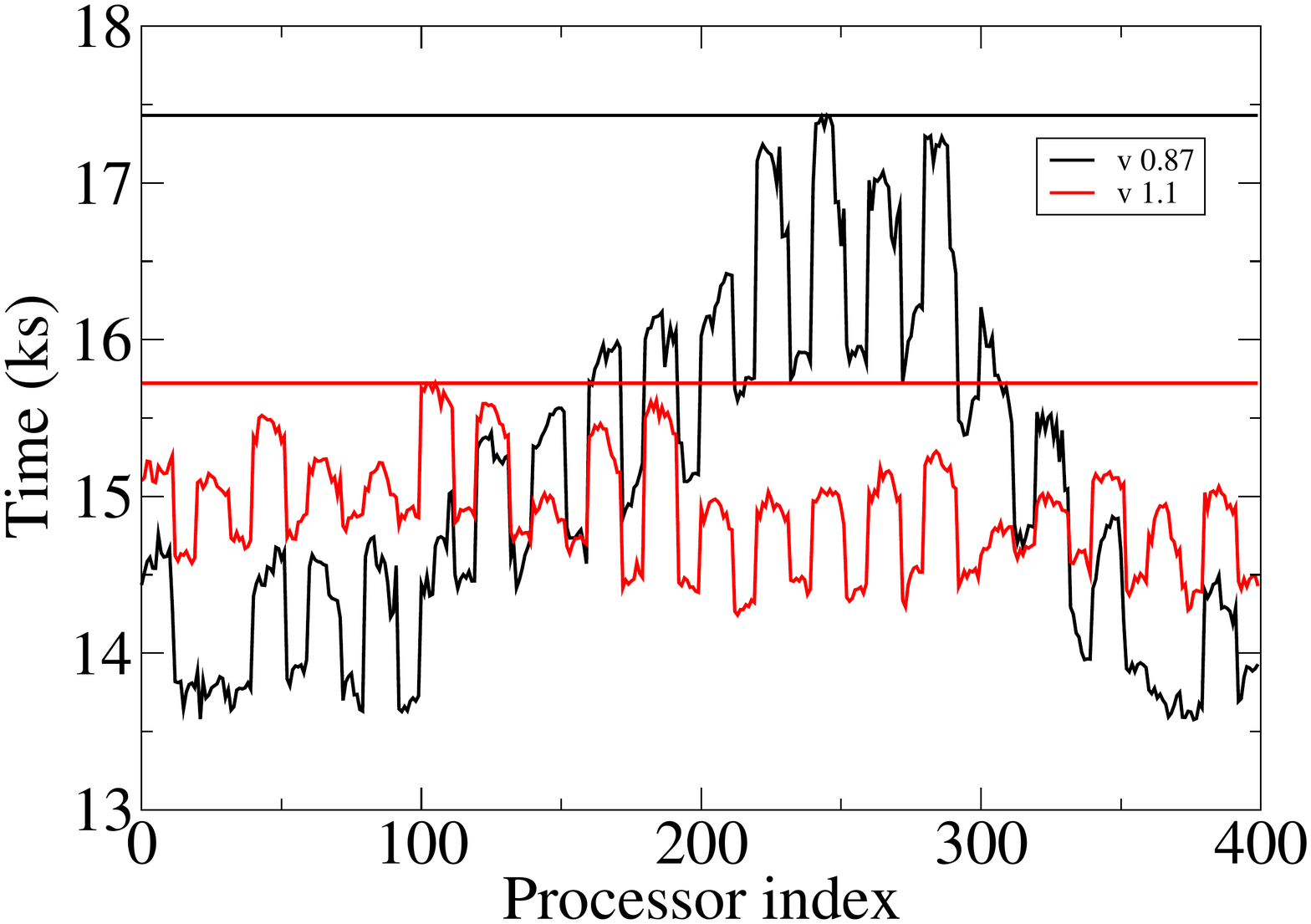}
  \caption{Total time of each processor $\mathrm{Ni^{3+}}$.
    The horizontal lines mark the total calculation time (maximum).}
  \label{fig:timeprocNi3}
\end{figure}


\begin{figure}
  \includegraphics[width=\textwidth]{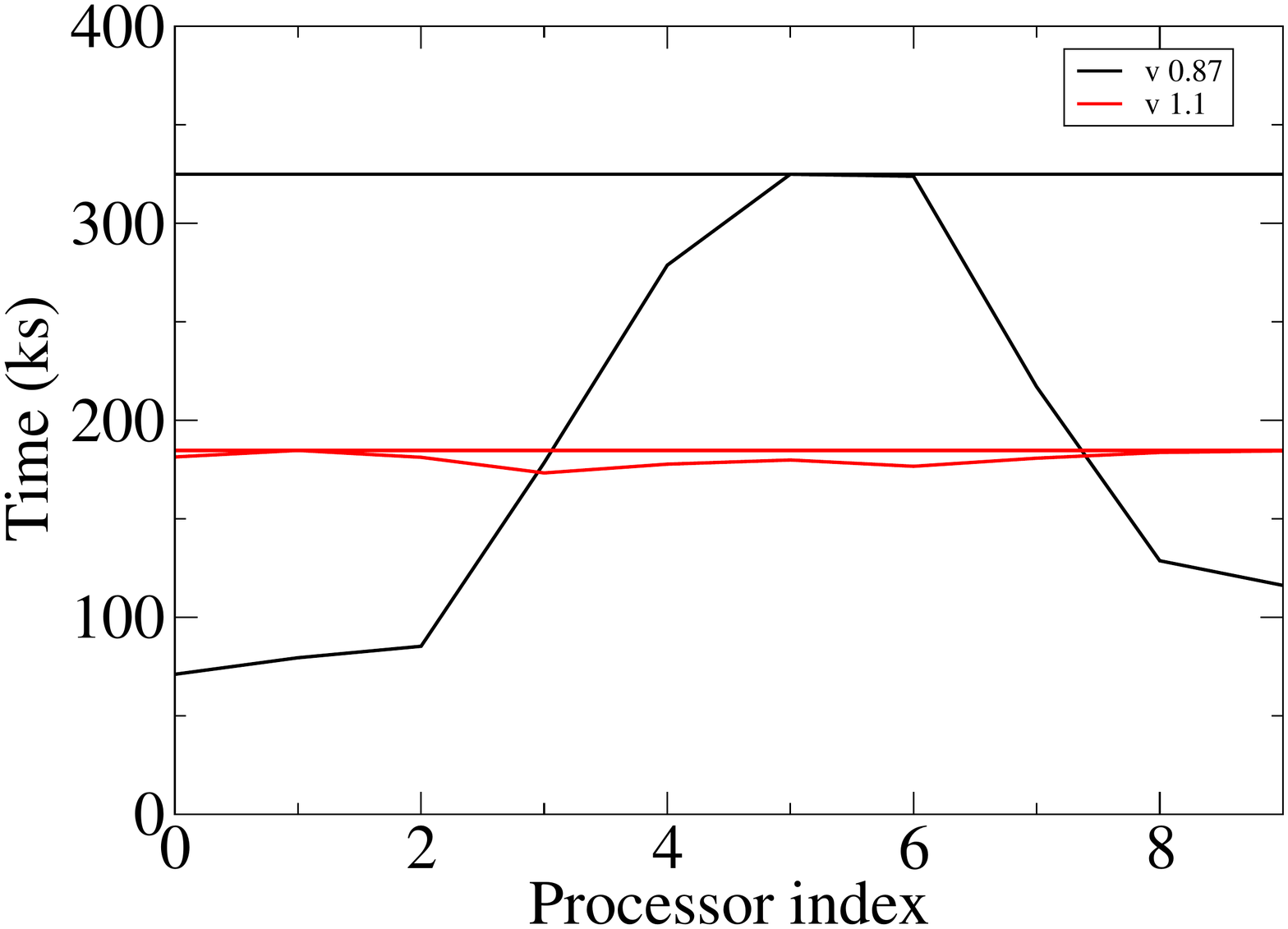}
  \caption{Total time of each processor $\mathrm{Fe}$ neutral.
    The horizontal lines mark the total calculation time (maximum).}
  \label{fig:timeprocFe1}
\end{figure}

We have used v1.1 to calculate the electron-impact excitation of the 
ion $\mathrm{Ni}^{3+}$,
for 42 $J \pi$ partial waves, with a maximum of 1818 channels per partial wave
and 10 energies per processor.
We compare the computation time with the version v0.87.
Figure \ref{fig:timeenNi3} shows the real nature of the problem, 
how the calculation time is very different for each value of the impact energy.
The difference between the named slow and fast energies is self-evident, reaching in 
the worst cases a factor ten.
The most of the calculation time will be spent in energies around the $150$ index.

Figure \ref{fig:timeprocNi3} illustrates the workload unbalance versus processor time.
When some processors have to calculate a larger number of slow energies, 
other processors do not. In the slowest cases, 
the relative difference in the processing time can reach a 20\%, 
this is unused processing time.
In contrast, twisting the distribution of energy points and forcing all processors to work
all energies, the workload balancing is greatly improved reducing the overall processing time
by approximately 15\%, and the maximum processor waiting time to less than one half.
Both curves in figure \ref{fig:timeprocNi3} have the same area,
but curve for v1.1 is flatter and its maximum (total calculation time) is
a 15\% smaller.

This difference is more evident in a more extreme calculation case.
Figure \ref{fig:timeprocFe1} shows a calculation case for neutral $\mathrm{Fe}$, 
with 10 partial waves and number of channels between 6943 and 7102.
In this case we perform the calculation in the most extreme case, 
one processor per energy, and equal number of partial waves than energies,
so each processor works only one symmetry in each energy.
The unequal workload balance is more extreme in this case,
reaching a comparative factor of 4 in the worst case.
The effect of the twisting of the energies and symmetries leads to a reduction of
a factor $1/2$ in the computing time.

\section{Example of input}
\label{sec:example}

We show an example the ASCII standard input ({\em dstgf}) for the case of
electron-impact excitation of ion $\mathrm{Ge}^{2+}$, 30-electron system plus
projectile.

For the inner region calculation, we carried out 42 relativistic partial waves, 
with $J=\frac{1}{2}$ to $20 \frac{1}{2}$, even and odd parity.
From the inner region,  42 {\em H.DATXXX} files were produced and 
one {\em sizeBP.dat} of length 42 lines, transcribing the partial wave 
symmetry associated with {\em H.DATXXX} file. 
section \ref{subsubsec:sizehdat}.

We calculate in the outer region the collision strengths for the first 40 partial
waves, from the first excitation threshold of the ion
($0.56 \Ry=0.14 z^2$) to twice the ionisation limit 
($5.0 \Ry=1.25 z^2$), with a fine energy mesh of $10^{-5} z^2$.

\noindent
\texttt{
 \&STGF IMESH=1 IQDT=0 IPRINT=-2 IOPT1=2 LRGLAM=-1 \&END \\
 \&MESH1 MXE=120000 E0=0.1300 EINCR=0.00001 \&END \\
 0 1 0 \\
 0 1 1 \\
 0 3 0 \\
 0 3 1 \\
 0 5 0 \\
 0 5 1 \\
 0 7 0 \\
 0 7 1 \\
 0 9 0 \\
 0 9 1 \\
 0 11 0 \\
 0 11 1 \\
 0 13 0 \\
 0 13 1 \\
 0 15 0 \\
 0 15 1 \\
 0 17 0 \\
 0 17 1 \\
 0 19 0 \\
 0 19 1 \\
 0 21 0 \\
 0 21 1 \\
 0 23 0 \\
 0 23 1 \\
 0 25 0 \\
 0 25 1 \\
 0 27 0 \\
 0 27 1 \\
 0 29 0 \\
 0 29 1 \\
 0 31 0 \\
 0 31 1 \\
 0 33 0 \\
 0 33 1 \\
 0 35 0 \\
 0 35 1 \\
 0 37 0 \\
 0 37 1 \\
 0 39 0 \\
 0 39 1 \\
 -1 -1 -1
}

With this input we will perform the calculation of the first 40 partial waves
specified in {\em sizeBP.dat} file and no top-up, 
with the full balancing efficiency of {\sc pstgf},
distributing read files in the different processors and doing the efficient 
energy split in processors and partial waves.

In order to use this efficient split, we are forced to deactivate the top-up
procedure, through setting $LRGLAM$ to negative.
As a second step we perform the high-$J$ top-up, 
in this case it is necessary that all processors calculate the collision
strengths for the same energies at least for the two last partial waves.
Hence, to perform the top-up, we carry out a second calculation including just
these two last partial waves.
In this second calculation, we do not take advantage of the efficiency improvements
of v1.1-2018, but it is a small calculation, so the loss in user time is minimal.

\noindent
\texttt{
 \&STGF IMESH=1 IQDT=0 IPRINT=-2 IOPT1=2 LRGLAM=43 \&END \\
 \&MESH1 MXE=120000 E0=0.1300 EINCR=0.00001 \&END \\
 0 41 0 \\
 0 41 1 \\
 -1 -1 -1
}

This time, we set the top-up variable $LRGLAM$ to the value of $2J$ of the first
partial wave to be topped-up $=41$.
With the simple post-processing tool {\sc omadd}, 
we can add-up the two $OMEGA$ files from the two calculations.

\section{Future work}
\label{sec:future}

All these improvements in the implementation have been applied to the undamped
version of {\sc pstgf}.
Nevertheless, they do not modify the physics of the problem.
As a consequence all these modifications can be integrated within the radiationally damped version 
{\sc pstgfdamp} \cite{whiteford2001} with a similar increase of efficiency.

Likewise, it would be necessary to modify the ICFT code {\sc pstgicf} \cite{griffin1998}
to allow for the reading of the files containing the $\mathbf{K}$ and $\mathbf{S}$ 
in a consistent way with current version of {\sc pstgf}.

\section*{Acknowledgments}

Present work has been funded by the STFC through the consolidated grant ST/P000312/1,
and the APAP grant.
The test calculations were carried out on the supercomputer ARCHER, 
property of the Engineering and Physical Science Research Council
under the allocations E464-RAMPA and E585-AMOR.

Finally, we acknowledge once more the first version of {\sc stgf}
written by Professor M J Seaton, a foundation for the current work 
of several people presented here.


\bibliographystyle{elsarticle-num}
\bibliography{references}

\begin{thebibliography}{10}
\expandafter\ifx\csname url\endcsname\relax
  \def\url#1{\texttt{#1}}\fi
\expandafter\ifx\csname urlprefix\endcsname\relax\def\urlprefix{URL }\fi
\expandafter\ifx\csname href\endcsname\relax
  \def\href#1#2{#2} \def\path#1{#1}\fi

\bibitem{seaton2002}
M.~Seaton, Coulomb functions for attractive and repulsive potentials and for
  positive and negative energies, Comp. Phys. Comm. 146~(2)
  (2002) 225--249.
\newblock \href {https://doi.org/10.1016/S0010-4655(02)00275-8}
  {\path{doi:10.1016/S0010-4655(02)00275-8}}.

\bibitem{seaton1983}
M.~J. Seaton, Quantum defect theory, Reports on Progress in Physics 46~(2)
  (1983) 167.
\newblock \href {https://doi.org/10.1088/0034-4885/46/2/002}
  {\path{doi:10.1088/0034-4885/46/2/002}}.

\bibitem{burke2011}
P.~G. Burke, R-Matrix of Atomic Collisions: Application to Atomic, Molecular,
  and Optical Processes, Springer-Verlag, New-York, 2011.

\bibitem{burke1992a}
V.~Burke, P.~Burke, N.~Scott, A new no-exchange R-matrix program, Comp. Phys.
  Comm. 69~(1) (1992) 76--98.
\newblock \href {https://doi.org/10.1016/0010-4655(92)90131-H}
  {\path{doi:10.1016/0010-4655(92)90131-H}}.

\bibitem{berrington1995}
K.~A. Berrington, W.~B. Eissner, P.~H. Norrington, RMATRX1: Belfast atomic
  R-matrix codes, Comp. Phys. Comm. 92 (1995) 290.

\bibitem{norrington1981}
P.~H. Norrington, I.~P. Grant, Electron scattering from Ne~{\sc ii} using the
  relativistic R-matrix method, J. Phys. B: At. Mol. Phys. 14 (1981)
  L261--L267.

\bibitem{norrington1987}
P.~H. Norrington, I.~P. Grant, Low-energy electron scattering by Fe~{\sc xxiii} and
  Fe~{\sc vii} using the Dirac R-matrix method, J. Phys. B: At. Mol. Opt. Phys. 20
  (1987) 4869--4881.

\bibitem{badnell1999c}
N.~R. Badnell, A perturbative approach to the coupled outer-region equations
  for the electron-impact excitation of neutral atoms, J. Phys. B:
  At., Mol. Opt. Phys. 32~(23) (1999) 5583.
\newblock \href {https://doi.org/10.1088/0953-4075/32/23/312}
  {\path{doi:10.1088/0953-4075/32/23/312}}.

\bibitem{griffin1998}
D.~C. Griffin, N.~R. Badnell, M.~S. Pindzola, R-matrix electron-impact
  excitation cross sections in intermediate coupling: an MQDT transformation
  approach, J. Phys. B: At. Mol. Opt. Phys. 31~(16) (1998) 3713--3727.
\newblock \href {https://doi.org/10.1088/0953-4075/31/16/022}
  {\path{doi:10.1088/0953-4075/31/16/022}}.

\bibitem{gorczyca1995}
T.~W. Gorczyca, M.~S. Pindzola, F.~S. Shieh, C.~L. McCreary, Adaptation of
  asymptotic close-coupling methods to massively parallel computers, Comp.
  Phys. Comm. 88~(2) (1995) 211.
\newblock \href {https://doi.org/10.1016/0010-4655(95)00068-Q}
  {\path{doi:10.1016/0010-4655(95)00068-Q}}.

\bibitem{badnell2011b}
N.~R. Badnell, A Breit-Pauli distorted wave implementation for autostructure,
  Comp. Phys. Comm. 182~(7) (2011) 1528--1535.
\newblock \href {https://doi.org/10.1016/j.cpc.2011.03.023}
  {\path{doi:10.1016/j.cpc.2011.03.023}}.

\bibitem{froese-fischer2007}
C.~Froese-Fischer, G.~Tachiev, G.~Gaigalas, M.~R. Godefroid, An MCHF
  atomic-structure package for large-scale calculations, Comp. Phys. Comm.
  176~(8) (2007) 559 -- 579.
\newblock \href {https://doi.org/10.1016/j.cpc.2007.01.006}
  {\path{doi:10.1016/j.cpc.2007.01.006}}.

\bibitem{dyal1989}
K.~G. Dyall, I.~P. Grant, C.~T. Johnson, F.~A. Parpia, E.~P. Plummer, GRASP: a
  general-purpose relativistic atomic structure program, Comp. Phys. Comm. 55
  (1989) 425.

\bibitem{parpia1996}
F.~A. Parpia, C.~F. Fischer, I.~P. Grant, GRASP92: a package for large-scale
  relativistic structure calculations, Comp. Phys. Comm. 94 (1996) 249.

\bibitem{hibbert1975}
A.~Hibbert, CIV3 - a general program to calculate configuration-interaction
  wave functions and electric-dipole oscillator strengths, Comp. Phys. Comm.
  9~(3) (1975) 141 -- 172.
\newblock \href {https://doi.org/10.1016/0010-4655(75)90103-4}
  {\path{doi:10.1016/0010-4655(75)90103-4}}.

\bibitem{burke1971}
P.~G. Burke, A.~Hibbert, W.~D. Robb, Electron scattering by complex atoms, J.
  Phys. B: At. Mol. Phys. 4~(2) (1971) 153.
\newblock \href {https://doi.org/10.1088/0022-3700/4/2/002}
  {\path{doi:10.1088/0022-3700/4/2/002}}.

\bibitem{berrington1987}
K.~A. Berrington, P.~G. Burke, K.~Butler, M.~J. Seaton, P.~J. Storey, K.~T.
  Taylor, Y.~Yan, Atomic data for opacity calculations. {\sc ii}. computational
  methods, J. Phys. B: At. Mol. Phys. 20~(23) (1987) 6379.
\newblock \href {https://doi.org/10.1088/0022-3700/20/23/027}
  {\path{doi:10.1088/0022-3700/20/23/027}}.

\bibitem{zatsarinny2006}
O.~Zatsarinny, BSR: B-spline atomic R-matrix codes, Comp. Phys. Comm. 174
  (2006) 273--356.
\newblock \href {https://doi.org/10.1016/j.cpc.2005.10.006}
  {\path{doi:10.1016/j.cpc.2005.10.006}}.

\bibitem{burgess1992}
A.~Burgess, J.~A. Tully, On the analysis of collision strengths and rate
  coefficients, Astron. Astroph. 254 (1992) 436--453.

\bibitem{whiteford2001}
A.~D. Whiteford, N.~R. Badnell, C.~P. Ballance, M.~G. O'Mullane, H.~P. Summers,
  A.~L. Thomas, A radiation-damped R-matrix approach to the electron-impact
  excitation of helium-like ions for diagnostic application to fusion and
  astrophysical plasmas, J. Phys. B 34~(15) (2001) 3179--3191.
\newblock \href {https://doi.org/10.1088/0953-4075/34/15/320}
  {\path{doi:10.1088/0953-4075/34/15/320}}.

\end{thebibliography}


\begin{thebibliography}{0}
\bibitem{1} M.~J. Seaton, Coulomb functions for attractive and repulsive potentials 
   and for positive and negative energies, Comp. Phys. Comm. 146 (2) (2002) 225--249. 
   doi:https://doi.org/10.1016/S0010-4655(02)00275-8.
\bibitem{2} M.~J. Seaton, Quantum defect theory, Rep. Prog. Phys. 46 (2) (1983) 167. 
   doi:https://doi.org/10.1088/0034-4885/46/2/002.
\end{thebibliography}


\end{document}